\documentclass[a4paper,11pt]{article}
\usepackage{amsfonts}
\usepackage{amsmath}
\usepackage{amssymb}
\usepackage{graphicx}
\usepackage{makeidx}
\usepackage{enumerate}
\usepackage{verbatim,color}

\makeindex

\def \be {\begin{equation}}

\def \ee {\end{equation}}
\def \bea {\begin{eqnarray}}
\def \eea {\end{eqnarray}}
\def \nn {\nonumber}

\def \rr {\raise.35ex\hbox{\small $\prime$}\kern-.17em{\mbox{\large $\imath$}}}

\def \dels {\partial\kern-.6em /\kern.1em}
\def \As {{A\kern-.5em / \kern.5em}}
\def \Ds {D\kern-.7em / \kern.5em}

\def \dag {\dagger}

\def \ks {k\kern-.5em /}

\newcommand{\mathsym}[1]{{}}

\newcommand{\ba}{\begin{eqnarray}}
\newcommand{\ea}{\end{eqnarray}}

\def\bfone{\relax{\rm 1\kern-.35em 1}}
 
\makeatletter \@addtoreset{equation}{section} \makeatother

 \allowdisplaybreaks[1]

\begin{document}

\begin{titlepage}
 \thispagestyle{empty}
 \begin{flushright}
     \hfill{CERN-PH-TH/2012-276}\\
	\hfill{DFPD-12/TH/13}\\
 \end{flushright}

 \vspace{50pt}

 \begin{center}
     { \Huge{\bf      {$d$-Geometries Revisited}}}

     \vspace{50pt}

     {\Large {\bf Anna Ceresole$^{a}$, \bf Sergio Ferrara$^{b,c}$,\\ \bf Alessandra Gnecchi$^{d}$\footnote{Affiliation as of October 1st, 2012:
     Institute for Theoretical Physics and Spinoza Institute, Utrecht University, 3508 TD Utrecht, The Netherlands.
  }  \bf and \bf Alessio Marrani$^{b}$}}

     \vspace{30pt}
  {\it ${}^a$ INFN - Sezione di Torino,\\
     Via P. Giuria 1, I-10125 Torino, Italy\\\texttt{ceresole@to.infn.it}}

\vspace{20pt}

      {\it ${}^b$ Physics Department, Theory Unit, CERN,\\
     CH -1211, Geneva 23, Switzerland\\\texttt{sergio.ferrara@cern.ch}\\\texttt{alessio.marrani@cern.ch}}

     \vspace{20pt}

   {\it ${}^c$ INFN - Laboratori Nazionali di Frascati,\\
     Via Enrico Fermi 40, I-00044 Frascati, Italy}
    \vspace{20pt}


   {\it ${}^d$ Dipartimento di Fisica, Universit\`{a} di Padova\\
     Via Marzolo 8, I-35131 Padova, Italy\\\texttt{alessandra.gnecchi@pd.infn.it}}

    \vspace{20pt}

     \vspace{50pt}

     {ABSTRACT}
 \end{center}

 \vspace{10pt}
\noindent
We analyze some properties of the four dimensional supergravity theories which originate from five dimensions upon reduction. They generalize to $N > 2$ extended supersymmetries the d-geometries with cubic prepotentials, familiar from N=2 special K\"ahler geometry. We emphasize the role of a suitable parametrization of the scalar fields and the corresponding triangular symplectic basis. We also consider applications to the first order flow equations for non-BPS extremal black holes.
 \vfill

\end{titlepage}

\baselineskip 6 mm



\section{\label{Intro}Introduction}

The allowed scalar manifolds for the $N=2$ five-dimensional supergravity
coupled to $n_{V}-1$ Abelian vector multiplets, parametrized by scalar
fields $\varphi ^{x}$ ($x=1,\ldots ,n_{V}-1$), can be described as the $%
\left( n_{V}-1\right) $-dimensional cubic hypersurface $\frac{1}{3!}d_{ijk}%
\hat{\lambda}^{i}\hat{\lambda}^{j}\hat{\lambda}^{k}=1$ of an ambient space
spanned by $n_{V}$ coordinates $\hat{\lambda}^{i}=\hat{\lambda}^{i}(\varphi
^{x})$ ($i=1,...,n_{V}$) \cite{GST-2}. The cubic nature of this polynomial
constraint is related to the presence of the Chern-Simons term $%
d_{ijk}F^{i}F^{j}A^{k}$ in the Lagrangian for the $n_{V}$ vector fields $%
A_{\mu }^{i}$ ($\mu =0,1,2,3,4$), with $n_{V}$ denoting the total number of $%
D=5$ vector potentials (including the $D=5$ graviphoton). A complete
classification of the allowed homogeneous scalar manifolds was given in \cite%
{deWit:1991nm,dWVVP}, and many interesting properties, especially when they
are restricted to be a symmetric coset of the Jordan family, were already
analyzed in \cite{GST-2}. When this theory is dimensionally reduced to four
dimensions, it yields a particular class of $N=2$ four-dimensional matter
coupled models with special K\"{a}hler target space geometries, which were
studied in \cite{dWVVP} under the name \textquotedblleft $d$-spaces".
There, the uplift between four and five dimensions was called
\textquotedblleft $r$-map", since it associates real scalars to the $N=2$
four dimensional complex scalar fields belonging to the $n_{V}$ $D=4$ vector
multiplets: $z^{i}=X^{i}/X^{0}=a^{i}-i~\lambda ^{i}$, with $a^{i},\lambda
^{i}$ real and with the index $0$ pertaining to the $D=4$ graviphoton. The
axions $a^{i}$ originate by Kaluza-Klein (KK) reduction from the vector
components $A_{4}^{i}$, and the $\lambda ^{i}=\hat{\lambda}^{i}e^{2\phi }$
are $n_{V}$ real scalars parametrizing the $D=5$ scalars $\phi ^{x}$ and the
KK scalar $\phi =g_{44}$. In this sense, the $r$-map is similar to the $c$%
-map, relating the moduli spaces of special K\"{a}hler vector multiplets to
the quaternionic hypermultiplets scalar manifolds in $N=2$ theories \cite%
{CFG,dWVVP}. In superstring theories, the c-map relates $IIA$ and $IIB$
string theories compactified on the same $(2,2)$ superconformal field theory
at $c=9$, while in a purely supergravity context, it can simply be viewed as
a consequence of dimensional reduction from $4$ to $3$ dimensions \cite{CFG}%
. Actually, these $N=2$ matter coupled theories, where the holomorphic
prepotential takes the cubic form
\begin{equation}
F\left( X\right) \equiv \frac{1}{3!}d_{ijk}\frac{X^{i}X^{j}X^{k}}{X^{0}}\,,
\label{FF}
\end{equation}%
were first studied in \cite{Cremmer:1984hj}, where they were shown to lead
to supergravity couplings with flat potentials characterized by the
completely symmetric rank-$3$ tensor $d_{ijk}$. They are particularly
relevant in connection with the large volume limit of Calabi-Yau
compactifications of type $IIA$ superstrings where the d-tensors are related
to intersection forms of the Calabi-Yau manifold.

\noindent Formally, the $d$-tensor appears in the expression for the
curvature of any special K\"{a}hler manifold \cite{CVP}
\begin{equation}
R_{i\bar{\jmath}k\bar{l}}=-g_{i\bar{\jmath}}g_{k\bar{l}}-g_{i\bar{l}}g_{k%
\bar{\jmath}}+C_{ikp}\overline{C}_{\bar{\jmath}\bar{l}\bar{p}}g^{p\bar{p}}
\label{FF2}
\end{equation}%
since in \textquotedblleft special coordinates" the covariantly holomorphic
quantity $C_{ijk}$ is given by $C_{ijk}=e^{K(z,\bar{z})}d_{ijk}$, with $K(z,%
\bar{z})$ denoting the K\"{a}hler potential.

\noindent Notice that a generic $d$-geometry of complex dimension $n_{V}$ is
not necessarily a coset space, but nevertheless it admits $n_{V}+1$ real
isometries, corresponding to Peccei-Quinn shifts of the $n_{V}$ axions, and
to an overall rescaling of the prepotential \cite{dWVVP}.

This paper aims to study $d$-geometries in a framework broader than $N=2$,
considering the $r$-map for $N\geq 2$ extended supergravities along the
lines of previous work on this 4D/5D relation in the context of black hole
supergravity solutions and their attractors \cite%
{CFM1,Ceresole:2009jc,Ceresole:2009id}. Due to the structure of 5D spinors,
these generalized $d$-geometries encompass all extended supergravities with
a number of supercharges multiple of $8$, and thus an {even} number of
supersymmetries $N=2,4,6,8$.

$d_{ijk}$ is an invariant tensor of the underlying classical duality group $%
G_{5}$ of the $D=5$ action \cite{Gaillard:1981rj}, corresponding to the
continuous version of the non-perturbative string symmetries $G_{5}(\mathbb{Z%
})$ of \cite{HT-1}. The dimensional reduction yields interesting relations
between the scalar manifolds and the isometries of the 5D and 4D theories: $%
G_{5}$ is embedded into the $D=4$ electric--magnetic duality group $G_{4}$,
whose isometries are included in $Sp(2n_{V}+2,\mathbb{R})$ (for generic $N>1$%
, one has $Sp(2n,\mathbb{R})$ for a theory with $n$ vector potentials; for $%
N=2$, $n=n_{V}+1$). More precisely, one always has the chain of embeddings
\begin{equation}
G_{5}\times SO(1,1)\subset G_{4}\subset Sp(2n_{V}+2,\mathbb{R}).
\end{equation}

Our main point is that the five-dimensional origin of all generalized
d-geometries naturally selects a particular branching of the $D=4$ scalars,
given by the axions $a^I$, the Kaluza-Klein scalar $\phi$ and the 5D scalars
$\lambda ^{x}$:
\begin{equation}
\Phi =\left\{ a^{I},\phi ,~\lambda^{x}\right\} .  \label{scalars-decomp}
\end{equation}%
When $N>2$ these latter transform in a suitable representation of $H_{5}$,
the maximal compact subgroup of $G_{5}$, which depends on $N$ : for
instance, in $N=8$ there are $42$ of them, sitting in the rank-$4$
antisymmetric skew-traceless representation $\mathbf{42}$ of $USp(8)$, and
there are $27$ axions.

Remarkably, only in $N=2$ the number of axions exactly matches the number of
scalars plus $1$, so that the two sets can be combined to give complex
scalars. For this case we will use a small index $i$ rather than $I$, to
emphasize its complex nature. We will illustrate that the $a^{I}$ and $\phi$
give rise to a universal sector which is present in any $N=2,4,6,8$
-extended supergravity in $D=4$ endowed with generalized $d$-geometry for
the vector multiplet sigma model.

In the study and classification of BPS and non-BPS extremal black hole
supergravity solutions, the relation between 4D and 5D for cubic holomorphic
prepotentials $F(X)$ (\ref{FF}) was used in \cite{CFM1} to relate the two $%
N=2$ effective black hole potentials and to derive the 4D attractors and
Bekenstein-Hawking classical entropies from the 5D ones. The key idea was to
reformulate the 4D effective black hole potential in terms of 5D real
special geometry data, implementing the natural splitting (\ref%
{scalars-decomp}) of the 4D scalar fields.

\noindent Some extra features arise in \textit{symmetric} special
geometries, where the $d$-symbols satisfy the relation \cite{GST-2}%
\begin{equation}
d_{r(pq}d_{ij)k}d^{rkl}=\frac{4}{3}\delta _{(p}^{l}d_{qij)},
\end{equation}%
and one can define cubic , $G_{5}$-invariant, and quartic, $G_{4}$-invariant
polynomials of electric ($q_{0},q_{i}$) and magnetic charges ($p^{0},p^{i}$)
by \cite{FG}:%
\begin{eqnarray}
I_{4}\left( p^{0},p^{i},q_{0},q_{i}\right) &=&-\left(
p^{0}q_{0}+p^{i}q_{i}\right) ^{2}+4\left[ q_{0}I_{3}\left( p\right)
-p^{0}I_{3}\left( q\right) +\frac{\partial I_{3}\left( q\right) }{\partial
q_{i}}\frac{\partial I_{3}\left( p\right) }{\partial p^{i}} \right] , \\
I_{3}(p) &\equiv &\frac{1}{3!}d_{ijk}p^{i}p^{j}p^{k},~I_{3}(q)\equiv \frac{1%
}{3!}d^{ijk}q_{i}q_{j}q_{k}\, .
\end{eqnarray}
The simplest example of rank-$3$ symmetric $d$-geometry is provided in $N=2$
by the $stu$ model \cite{STU}, with $3$ complex scalar fields spanning the
coset $(SU(1,1)/U(1))^{3}$, which serves as the ubiquitous toy model in the
context of black holes arising from superstring and $M$-theory.

The generalization of $N=2$ special geometry is achieved in terms of a
generalized symplectic formalism, established in \cite{Ferrara:2006em},
which enlarges the rich geometric structure of special K\"{a}hler manifolds
\cite{dWVVP} to the other extended supergravities. In fact, an important
difference between $N=2$ and $N>2$ extended theories is that for $N>2$ the
scalar sigma model is always given by a symmetric space $G/H$.

The formalism of \cite{Ferrara:2006em} hinges on the definition of
generalized sections $(\mathbf{f},\mathbf{h})$ of a flat symplectic bundle
\cite{ADF-revisited}, which relates to $N>2$ the flat bundle underlying
special K\"{a}hler geometry \cite{Strominger-SKG}. Even in $N=2$ the
sections are fundamental, since they allow to describe also theories where
the holomorphic prepotential $F(X^{\Lambda })$ does not exist \cite%
{Ceresole:1994cx}. More precisely, the sections $V_{A}=(f_{A}^{\Lambda },h_{{%
\Lambda }A})$, with $\Lambda =0,\ldots ,n_{V}$ and $A=0,a$, are square
complex matrices defined in $N=2$ supergravity by
\begin{equation}
(\mathbf{f},\mathbf{h})=(L^{\Lambda },\overline{D}_{\bar{a}}\overline{L}%
^{\Lambda };M_{\Lambda },\overline{D}_{\bar{a}}\overline{M}_{\Lambda })\,,
\label{usual}
\end{equation}%
with $(L^{\Lambda },M_{\Lambda })=e^{K/2}(X^{\Lambda },F_{\Lambda })$, $%
D_{a} $ denoting the flat covariant derivative in the scalar manifold: $%
D_{a}=e_{i}^{a}D_{i}$, $g_{i\bar{\jmath}}=e_{i}^{a}e_{\bar{\jmath}%
}^{b}\delta _{ab}$ and $D_{i}=\partial _{i}+\frac{1}{2}\partial _{i}K$. They
satisfy
\begin{equation}
h_{\Lambda A}=\mathcal{N}_{\Lambda \Sigma }f_{\phantom\Sigma A}^{\Sigma }
\end{equation}%
where $\mathcal{N}_{\Lambda \Sigma }(z)$ is the 4D complex vector kinetic
matrix. The sections encode a generic element $\mathbf{L}$ of the flat $%
Sp(2n_{V}+2,\mathbb{R})$-bundle over the $D=4$ scalar manifold as \cite%
{Ferrara:2006em}
\begin{equation}
\left(
\begin{array}{cc}
A & B \\
C & D%
\end{array}%
\right) \longrightarrow \left(
\begin{array}{c}
\mathbf{f} \\
\mathbf{h}%
\end{array}%
\right) =\frac{1}{\sqrt{2}}\left(
\begin{array}{cc}
A & -\mathrm{i}B \\
C & -\mathrm{i}D%
\end{array}%
\right) \ ,  \label{deff1}
\end{equation}%
or the inverse transformation
\begin{equation}
\mathbf{L}\equiv \left(
\begin{array}{cc}
A & B \\
C & D%
\end{array}%
\right) =\sqrt{2}\left(
\begin{array}{cc}
\mathrm{Re}\,\mathbf{f} & -\mathrm{Im}\,\mathbf{f} \\
\mathrm{Re}\,\mathbf{h} & -\mathrm{Im}\,\mathbf{h}%
\end{array}%
\right) \ ,  \label{deff}
\end{equation}
with the symplectic property $\mathbf{L}^T\Omega \mathbf{L}=\Omega=\left(^{0-%
\mathbf{1}}_{{\mathbf{1}}{\phantom -}0} \right)$ yielding the conditions
\begin{equation}
i({\mathbf{f}}^{\dag }{\mathbf{h}}-{\mathbf{h}}^{\dag }{\mathbf{f}})=%
\relax{\rm 1\kern-.35em 1}\,,\qquad {\mathbf{f}}^{T}{\mathbf{h}}-{\mathbf{h}}%
^{T}{\mathbf{f}}=0\,.  \label{prop}
\end{equation}

This paper studies in detail the properties of a certain parametrization (%
\ref{partial-Iwa}), (\ref{L-fin}) of four-dimensional generalized $d$%
-geometries, which reflects their five-dimensional origin, yielding a
lower-triangular structure (\ref{L-fin}) for the matrix $\mathbf{L}$
characterizing the flat symplectic bundle sigma model which generalizes the
one of $N=2$ special K\"{a}hler $d$-geometry to any for any $N=2,4,6,8$.
This parametrization exploits nilpotent (of degree $4$) translations \cite%
{Strominger-SKG,CV,Lerche-1} parametrized by axion scalars $a^{I}$, and it
acts on the same space where the $d$-tensor is defined. The sigma model is
parametrized by additional block diagonal elements in the matrix $\mathbf{L}$%
, one of them being a dilatation in terms of the KK radius $\phi $, and by a
symmetric matrix, which depends on the $5D$ data and is related to the
kinetic term of the $5D$ vector fields.

It should be stressed that the proposed basis turns out to be different from
the standard parametrization of $N=2$ $d$-geometry (\ref{usual}), although
it leads to the same $4D$ vector kinetic matrix. We will emphasize that the
two symplectic frames are in fact related by a unitary transformation $M$
that was introduced in \cite{Ceresole:2009id}, which only depends on the $5D$
data. The unitary transformation $M$, that rotates the usual $N=2$ complex
basis of special geometry into the basis where $\mathbf{f}$ is real and $%
\mathbf{L}$ is lower triangular, allows to make a precise connection with
the ${N}=2$ $stu$ model, viewed as a sub sector of the full $N=8$ theory
\cite{Ferrara:2006em,Ceresole:2009iy,Ceresole:2009vp}. In the $t^{3}$ model,
this unitary transformation is numerical (\textit{cfr.} App. B), because the
relevant $5D$ uplifted theory is the pure $N=2$, $D=5$ supergravity.

Symmetric $d$-geometries can be related to Euclidean \textit{Jordan algebras}
of rank $3$ \cite{GST-2,Gunaydin:1983rk}, which were classified in \cite%
{JVNW}; in this case, the nilpotent axionic translations fit into a Jordan
algebra irreducible representation. The reduction to $D=4$ yields a \textit{%
Freudenthal triple system} (see \textit{e.g.} \cite{FG}).

Our results have interesting applications to non-BPS extremal black holes,
that we illustrate by making a precise and non trivial comparison between
the methods of \cite{Ceresole:2009iy} and \cite{Bossard:2009we} in the
computation of the \textsl{fake superpotential} \cite{Ceresole:2007wx} for
non-BPS solutions and $(p^{0},q_{0})$ charge configuration in the
stu-truncation of ${N}=8$ supergravity.

Beyond their interest in relation to supergravity structure and solutions,
one may hope that these general properties of $N\geq 2$ $d$-geometries and
the corresponding triangular symplectic frame (with degree-$4 $ nilpotent
axionic translations) could play a role in understanding the symmetry
structure of supergravity counterterms, in order to clarify the issue of
ultraviolet finiteness of $N=8$ and other extended supergravity theories in $%
D=4$ space-time dimensions \cite{K-private}.\medskip \medskip

The paper starts in Sec. \ref{sectADG} with the universal decomposition for
the $D=4$ symplectic element $\mathbf{L}$ in the proposed basis \ref%
{scalars-decomp}, where axion are singled out. Then, the relation between $%
\mathbf{L}$ and the matrix $\mathcal{M}$ entering the black hole effective
potential is elucidated in\ Sec. \ref{M-L}. Other geometrical identities in
a $5$-dimensionally covariant formalism are presented in Sec. \ref{Ids-5D}.
The simpler case of $N=4$, $D=4$ pure supergravity (with no matter coupling)
is discussed in\ Sec. \ref{N=4-D=4}. For $d$-geometries based on symmetric
spaces $G/H$, the computation of the \textit{Vielbein} and of the $H$%
-connection is carried out in Sec. \ref{Symmetric}, in particular focusing
on $N=8$ supergravity. Next, in Sec. \ref{Flat} the $N=2$ axion basis is
related to the reformulation of special K\"{a}hler geometry as flatness
condition of a symplectic connection \cite{Strominger-SKG}.

A detailed treatment of $N=2$ $d$-geometries is then given in\ Sec. \ref{N=2}%
, where we elaborate on the results of \cite{Ceresole:2009id} on the unitary
matrix $M$ rotating the axion basis to the usual special coordinates one.
Geometrical identities for $M$ and the related matrix $\widehat{M}$ are
derived in Sec. \ref{Sec-2}.

An application of the axion basis to the first order formalism for extremal
black holes is considered in\ Sec. \ref{FO}. After a preliminary analysis
for the $stu$ model in Secs. \ref{D0-D6} and \ref{U(1)^3} , explicit
computations for the $t^{3}$ limit in the $\left( p^{0},q_{0}\right) $ ($%
D0-D6$) charge configuration are performed in Secs. \ref{T^3-Model}, and the
known fake non-BPS superpotential is retrieved in Sec. \ref{Recover}. In
Table 1 we list the allowed Rank-$3$ Euclidean Jordan algebras $J_{3}$ and
corresponding symmetric generalized $d$-geometries, characterized by a
parameter $q$ related to the number of vector and scalar fields for each $%
N=2,4,6,8$.

Some appendices conclude the paper. In App. \ref{Res-Exp} useful results on
exponential matrices are collected, while App. \ref{Expl-T^3} contains some
explicit computations in the $t^{3}$ model, displaying the matrix $M$. The
purely imaginary nature of the \textit{Vielbein} of the $stu$ model and its
consistent embedding into the $N=8$ theory are discussed in App. \ref{App-1}%
. Finally, App. \ref{App-2} deals with the duality-invariant polynomial and
the first order  fake superpotential in the $D0-D6$ configuration of the $stu
$ model with $i_{3}=0$.

\newpage

\section{\label{sectADG}Universal Decomposition for the $D=4$ Symplectic
Element in the Axion Basis}

We are interested in general features of all $D=4$ Maxwell-Einstein
(super)gravity theories admitting an uplift to $D=5$.
The classification of the tensors $d_{IJK}$ associated to homogeneous Riemannian $d$%
-spaces was performed in \cite{dWVVP}.
For symmetric geometries, $d_{IJK}$ can be characterized as the
cubic norm of an associated rank-$3$ Jordan algebra \footnote{%
With the exception of the \textit{non-Jordan symmetric sequence} \cite%
{deWit:1992cr} of $N=2$, $D=5$ vector multiplets' scalar manifolds $\frac{%
SO(1,n_{V})}{SO(n_{V})}$.} \cite{JVNW,GST-2}.
 In this case, the general properties are given in terms of a parameter $q$
reported in Table 1. 

The number of $D=5$ vectors is $n_{V}=3q+3$, while the
number of $D=4$ $2$-form field strengths and their duals is $6q+8$. Only in $%
N=2$ theories, the number of 5D real scalars is $3q+2$, while the number of
4D complex scalars is $3q+3$ (one for each 4D Abelian vector multiplet).
Quite generally, the relation between the number of vector and scalar fields
in theories derived from five dimensions is such that
\begin{eqnarray}
\#~4D\text{ scalars~} &=&\#~5D\text{ scalars}+\#~5D\text{ vectors}+1\,
\notag \\
\#~4D\text{ vectors~} &=&\#~5D\text{ vectors}+1=n_{V}+1\,,
\end{eqnarray}%
where the $n_{V}$ axions arise from the total number of $5D$ vectors.

We will show that in these generalized $d$-geometries, the representation of
the $D=4$ axions $a^{I}$ is nilpotent of degree four and that, together with
the Kaluza-Klein $SO(1,1)$ radius parametrized by the real scalar $\phi $,
it provides a universal sector of the scalar manifold of the $D=4$ theory,
regardless of its specific geometry. This reflects the property of special K%
\"{a}hler $d$-geometries \cite{dWVVP}, of always having as minimal isometry
of the scalar manifold the $n_{V}$ axionic Peccei-Quinn translations and the
$SO(1,1)$ overall rescaling.\medskip

To prove the above statement, we split the symplectic element $\mathbf{L}$
according to the decomposition of the $D=4$ scalars (\ref{scalars-decomp}),
and we demonstrate that\footnote{%
In the following we will switch the axion index from $i$ into $I$, whenever
our analysis holds for generic $N\geqslant 2$ $d$-geometries.}
\begin{equation}
\mathbf{L}\left( a^{I},\phi ,E\left( \lambda \right) \right) =\mathcal{A}%
(a^{I})\mathcal{D}(\phi )\mathcal{G}(E)\ .  \label{partial-Iwa}
\end{equation}%
In order to identify the various factors in (\ref{partial-Iwa}), one must
consider the definition (\ref{deff}) and complement it with the results of
\cite{Ceresole:2009id}, where the 4D/5D connection was used for $N=8$ to
determine the $28\times 28$ symplectic sections $(f_{A}^{\Lambda
},h_{\Lambda A})$ in a five-dimensionally covariant symplectic frame, where
the indices split as $\Lambda =(0,I)$ and $A=(0,a)$. They take the form:

\begin{equation}
f_{\phantom{\Lambda}A}^{\Lambda }=\frac{1}{\sqrt{2}}\left(
\begin{array}{c|c}
&  \\
e^{-3\phi }\  & \ 0 \\
&  \\ \hline
&  \\
\vspace{-1pt}e^{-3\phi }a^{I}\  & e^{-\phi }(a^{-1/2})_{\phantom Ia}^{I}%
\end{array}%
\right) \ ;  \label{2.21}
\end{equation}%
\begin{equation}
h_{\Lambda \,A}=\frac{1}{\sqrt{2}}\left(
\begin{array}{c|c}
&  \\
-e^{-3\phi }\frac{d}{6}-ie^{3\phi }\  & \ -\frac{1}{2}e^{-\phi
}d_{K}(a^{-1/2})_{\phantom Ia}^{K}+ie^{\phi }a^{K}(a^{1/2})_{K}^{\phantom Ka}
\\
&  \\ \hline
&  \\
\vspace{-1pt}\frac{1}{2}e^{-3\phi }d_{I}\  & e^{-\phi }d_{IJ}(a^{-1/2})_{%
\phantom Ja}^{J}-ie^{\phi }(a^{1/2})_{I}^{\phantom Ia}%
\end{array}%
\right) \ ,  \label{2.22}
\end{equation}%
with%
\begin{equation}
d\equiv d_{IJK}a^{I}a^{J}a^{k}\ ,\qquad d_{I}\equiv d_{IJK}a^{J}a^{k}\
,\qquad d_{IJ}\equiv d_{IJK}a^{K}\ ,
\end{equation}%
and where
\begin{equation}
E\left( \lambda \right) \equiv (a^{-1/2})_{a}^{~J}=E_{a}^{~J}  \label{rell-1}
\end{equation}%
is the coset representative of the 5D scalar manifold $G_{5}/H_{5}$. Notice
that in this basis the section $\mathbf{f}$ is real and it takes a lower
triangular form, and that the 5D scalars enter the sections only through $%
E(\lambda )$.

By generalizing this 5D/4D approach to the class of theories under
consideration and interpreting the indices $\Lambda ,A$ on the appropriate
representations, we determine the generic expression for each factor in (\ref%
{partial-Iwa}).

\noindent The axionic generators
\begin{equation}
\mathcal{A}(a)\equiv e^{T(a)}\ ,  \label{def-1}
\end{equation}%
also appeared in \cite{ADFL-flat} in the context of gauging of flat groups
in 4D supergravity, and they are given by the $2(n_{V}+1)\times2(n_{V}+1)$
block-matrix
\begin{equation}
T(a)=\left(
\begin{array}{cc|cc}
0 & 0 & \ 0 & 0 \\
a^{J} & 0 & \ 0 & 0 \\ \hline
0 & 0 & \ 0 & -a^{I} \\
0 & d_{IJ} & \ 0 & 0%
\end{array}%
\right) .  \label{T}
\end{equation}%
It is easily checked that $T(a)$ is nilpotent of order four:
\begin{equation}
T^{4}(a)=0\Rightarrow \mathcal{A}(a)=\relax{\rm 1\kern-.35em 1}+T(a)+\frac{1%
}{2}T^{2}(a)+\frac{1}{3!}T^{3}(a),
\end{equation}%
%
which, by definition (\ref{def-1}), yields
\begin{equation}
\mathcal{A}(a)=\left(
\begin{array}{cc|cc}
1 & 0 & \ 0 & 0 \\
a^{J} & 1 & \ 0 & 0 \\ \hline
-\frac{1}{6}d & -\frac{1}{2}d_{I} & \ 1 & -a^{I} \\
\frac{1}{2}d_{J} & d_{IJ} & \ 0 & 1%
\end{array}%
\right) \ .  \label{A}
\end{equation}%
As we will discuss in Sec. \ref{N=2}, this is in agreement with the $N=2$
interpretation of \cite{Lerche-1}. The $1$-dimensional Abelian $SO(1,1)$
factor in (\ref{partial-Iwa}) is given by
\begin{equation}
\mathcal{D}(\phi )=\left(
\begin{array}{cc|cc}
e^{-3\phi } & 0 & \ 0 & 0 \\
0 & e^{-\phi } & \ 0 & 0 \\ \hline
0 & 0 & \ e^{3\phi } & 0 \\
0 & 0 & \ 0 & e^{\phi }%
\end{array}%
\right) \ ,  \label{D}
\end{equation}%
whereas the $\left( 2n_{V}+2\right) \times \left( 2n_{V}+2\right) $ matrix $%
\mathcal{G }$ is
\begin{equation}
\mathcal{G}(\lambda )=\left(
\begin{array}{cc|cc}
1 & 0 & \ 0 & 0 \\
0 & E & \ 0 & 0 \\ \hline
0 & 0 & \ 1 & 0 \\
0 & 0 & \ 0 & E^{-1}%
\end{array}%
\right) \ .  \label{G}
\end{equation}%
%
%
By matrix multiplication of (\ref{A})-(\ref{G}) according to (\ref%
{partial-Iwa}), one finds that the symplectic matrix $\mathbf{L}$ (\ref{deff}%
) acquires the triangular form:
\begin{equation}
\mathbf{L}(a^{I},\phi ,E\left( \lambda \right) )\mathbf{=}\left(
\begin{array}{cc|cc}
e^{-3\phi } & 0 & \ 0 & 0 \\
a^{I}e^{-3\phi } & E_{a}^{~I}e^{-\phi } & \ 0 & 0 \\ \hline
-\frac{1}{6}de^{-3\phi } & -\frac{1}{2}d_{K}E_{{\phantom K}a}^{K}e^{-\phi }
& \ e^{3\phi } & -a^{K}(E^{-1})_{~K}^{a}e^{\phi } \\
\frac{1}{2}d_{I}e^{-3\phi } & d_{IK}E_{a}^{~K}e^{-\phi } & \ 0 &
(E^{-1})_{~I}^{a}e^{\phi }%
\end{array}%
\right) \ .  \label{L-fin}
\end{equation}%
We see that, in this particular basis, $B=\mathrm{Im}\,\mathbf{f}=0$, since
the $\mathbf{f}$ section is purely real:
\begin{equation}
\mathbf{f}=\mathrm{Re}\mathbf{f}=\frac{1}{\sqrt{2}}A(a^{I},\phi ,E\left(
\lambda \right) ).  \label{L-fin-2}
\end{equation}%
On the other hand, one has
\begin{equation}
\mathbf{h}=\frac{1}{\sqrt{2}}(C-iD)\quad \Rightarrow \quad
\begin{array}{l}
\mathrm{Re}\mathbf{h}=\frac{1}{\sqrt{2}}C(a^{I},\phi ,E\left( \lambda
\right) ,d_{IJK}) \\
\mathrm{Im}\mathbf{h}=-\frac{1}{\sqrt{2}}D(a^{I},\phi ,E\left( \lambda
\right) )\ ,\notag\label{L-fin-3}%
\end{array}%
\end{equation}%
along with the normalization
\begin{equation}
\mathbf{f}^{T}\mathrm{Im}\mathbf{h}=\frac{1}{2}.
\end{equation}%
Notice that the $C$ sub-block is the only one depending on $d_{IJK}$.

Conversely, one can say that the formula (\ref{L-fin}) for the symplectic
representative yields an explicit expressions for the symplectic sections $%
\mathbf{f}$\textbf{\ }and $\mathbf{h}$ which match Eqs. (\ref{2.21}) and (%
\ref{2.22}).

\medskip \medskip To make the discussion concrete, let us consider $N=8$
supergravity \cite{CJ-1}, based on the rank-$3$ Euclidean Jordan algebra $%
J_{3}^{\mathbb{O}_{s}}$ over the split octonions; the $D=5$ $U$-duality
group is $G_{5}=E_{6(6)}$ and $d_{IJK}$ is the invariant tensor of the
fundamental irrep. $\mathbf{27}$ ($I,J,K=1,...,27=n_{V}-1$, $x=1,...,42$, $%
i=1,...70$). The $Sp(56,\mathbb{R})$ matrix $\mathbf{L}$ (\ref{deff}) is the
coset representative of the rank-$7$ symmetric $D=4$ scalar manifold%
\begin{equation}
\frac{G_{4}}{H_{4}}=\frac{E_{7(7)}}{SU(8)},~\text{dim}_{\mathbb{R}}=70,
\label{D=4-coset}
\end{equation}%
where $H_{4}$ is the maximal compact subgroup of $E_{7(7)}$. The $70$ real $%
D=4$ scalars $z^{i}$ sit in the rank-$4$ self-real antisymmetric irrep. $%
\mathbf{70}$ of $SU(8)$.

The symplectic sections (\ref{2.21}) and (\ref{2.22}) are given in the
particular symplectic frame defined by the \textit{partial} decomposition of
$\mathbf{L}$ (\ref{L-fin}) in a \textit{solvable} basis, which is covariant
with respect to $H_{5}=USp(8)$, the local symmetry of the $D=5$ uplifted
theory. Furthermore, $E\left( \lambda \right) $ is the coset representative
of the rank-$6$ symmetric $D=5$ scalar manifold%
\begin{equation}
\frac{G_{5}}{H_{5}}=\frac{E_{6(6)}}{USp(8)},~\text{dim}_{\mathbb{R}}=42.
\label{D=5-coset}
\end{equation}%
The $42$ real $D=5$ scalars $\lambda ^{x}$ form the rank-$4$ self-real
antisymmetric skew-traceless irrep. $\mathbf{42}$ of $USp(8)$. Note that (%
\ref{rell-1}) is consistent with the well known fact that the $N=8$, $D=5$
kinetic vector matrix $(a^{-1})_{I}^{~J}$ is the square of the $D=5$ coset
representative \cite{ADF-revisited}. The scalar decomposition (\ref%
{scalars-decomp}) in this case becomes
\begin{eqnarray}
SU(8) &\supset &USp(8);  \notag  \label{dec-1} \\
\mathbf{70} &=&\underset{\lambda ^{x}}{\mathbf{42}}+\underset{a^{I}}{\mathbf{%
27}}+\underset{\phi }{\mathbf{1}},  \label{dec-2}
\end{eqnarray}%
where the axions $a^{I}$ form a representation of $J_{3}^{\mathbb{O}_{s}}$,
because%
\begin{eqnarray}
E_{6(6)} &\supset &USp(8);  \notag  \label{dec-3} \\
\mathbf{27} &=&\mathbf{27}.  \label{dec-4}
\end{eqnarray}

\section{\label{M-L}Relation between $\mathcal{M}$ and $\mathbf{L}$}

We now consider a further consequence of the symplectic structure of \textit{%
generalized special geometry} \cite{Ferrara:2006em}, holding for every $D=4$
Maxwell-Einstein supergravity even beyond d-geometries. It can be useful in
the present context and in view of applications to black holes. The black
hole effective potential for dyonic charges $Q=(p^\Lambda,q_\Lambda)$ is
given by \cite{Ferrara:1996dd}
\begin{equation}
V_{BH}=-\frac12 Q^t \mathcal{M}(\mathcal{N}) Q=<Q,V_A><Q,{\overline V}^A>
=Z_A\overline Z^A
\end{equation}
where the central charges $Z_A=<Q,V_A>$ are defined by the symplectic
product
\begin{equation}
Z_A=<Q,V_A>=Q^T\Omega V_A=f^\Lambda_{\ \
A}q_\Lambda-h_{\Lambda~A}p^\Lambda\, ,  \label{centralcharge}
\end{equation}
in terms of the symplectic invariant metric
\begin{equation}
\Omega =\left(
\begin{array}{cc}
0 & -\relax{\rm 1\kern-.35em 1} \\
\relax{\rm 1\kern-.35em 1} & 0%
\end{array}%
\right) .
\end{equation}

The matrix $\mathcal{M}$ is given by
\begin{eqnarray}
\mathcal{M} &=&\left(
\begin{array}{cc}
\relax{\rm 1\kern-.35em 1} & -\mathrm{Re}\mathcal{N} \\
0 & \relax{\rm 1\kern-.35em 1}%
\end{array}%
\right) \left(
\begin{array}{cc}
\mathrm{Im}\mathcal{N} & 0 \\
0 & \left( \mathrm{Im}\mathcal{N}\right) ^{-1}%
\end{array}%
\right) \left(
\begin{array}{cc}
\relax{\rm 1\kern-.35em 1} & 0 \\
-\mathrm{Re}\mathcal{N} & \relax{\rm 1\kern-.35em 1}%
\end{array}%
\right) \equiv \mathcal{R}^{T}\mathcal{M}_{D}\mathcal{R}\ ;  \label{M-call}
\\
\mathcal{R} &\equiv &\left(
\begin{array}{cc}
\relax{\rm 1\kern-.35em 1} & 0 \\
-\mathrm{Re}\mathcal{N} & \relax{\rm 1\kern-.35em 1}%
\end{array}%
\right) ;  \label{R} \\
\mathcal{M}_{D} &\equiv &\left(
\begin{array}{cc}
\mathrm{Im}\mathcal{N} & 0 \\
0 & \left( \mathrm{Im}\mathcal{N}\right) ^{-1}%
\end{array}%
\right) ,  \label{emme}
\end{eqnarray}%
where $\mathcal{N}=\mathbf{hf}^{-1}$ is the $D=4$ kinetic vector matrix.

In \textit{generalized special geometry } \cite{Ferrara:2006em} one
introduces the $Sp(2n_{V}+2)$ Hermitian matrix
\begin{equation}
\mathcal{C}\equiv \frac{1}{2}(\mathcal{M}+i\Omega )\ ;~~\mathcal{C}^{\dag }=%
\mathcal{C},
\end{equation}%
whose symmetric and antisymmetric parts are given by (\ref{M-call}) and $%
\Omega $ respectively. $\mathcal{C}$ is related to the symplectic sections $(%
\mathbf{f},\mathbf{h})$ by :
\begin{equation}
\mathcal{C}=\left(
\begin{array}{cc}
-\mathbf{hh}^{\dag } & \mathbf{hf}^{\dag } \\
\mathbf{fh}^{\dag } & -\mathbf{ff}^{\dag }%
\end{array}%
\right) \ ,
\end{equation}%
and therefore its action on the vector $V_{A}$ is given by
\begin{equation}
\frac{1}{2}(\mathcal{M}+i\Omega )V_{A}=i\Omega V_{A}\Leftrightarrow \mathcal{%
M}V_{A}=i\Omega V_{A},  \label{ress}
\end{equation}%
expressing a \textit{twisted self-duality} \cite{TSD}, recently used in \cite%
{BHen}.

Using the above relations, since both $\mathcal{M}$ and $\mathbf{L}$ are
given in terms of the sections $(\mathbf{f},\mathbf{h})$, one can see that
they can be related by \cite{Andrianopoli:2006ub,Aschieri:2008ns}
\begin{gather}
\mathcal{M}=-(\mathbf{L}^{T})^{-1}\mathbf{L}^{-1}=-(\mathbf{L}\,\mathbf{L}%
^{T})^{-1};  \label{defM1} \\
\Updownarrow  \notag \\
\mathcal{M}\mathbf{L}=-(\mathbf{L}^{T})^{-1}=\Omega \mathbf{L}\Omega ,
\label{defM2}
\end{gather}%
where the last step in (\ref{defM2}) follows from the symplecticity of $%
\mathbf{L}$ itself. Notice that, since also $\mathcal{M}$ is symplectic, (%
\ref{defM1}) implies that $\mathcal{M}=-\tilde{\mathbf{L}}\tilde{\mathbf{L}}%
^{T}$, with $\tilde{\mathbf{L}}\equiv \Omega \mathbf{L}$.

To prove (\ref{defM1})-(\ref{defM2}), one just notices that $\mathbf{L}$ (%
\ref{deff}) can be rewritten as (with $\ast $ here denoting complex
conjugation)
\begin{eqnarray}
\mathbf{L} &=&\frac{1}{\sqrt{2}}(\mathcal{B}+\mathcal{B}^{\ast }); \\
\mathcal{B} &\equiv &\left(
\begin{array}{cc}
\mathbf{f} & \ i\mathbf{f} \\
\mathbf{h} & \ i\mathbf{h}%
\end{array}%
\right) =\left(
\begin{array}{c}
\mathbf{f} \\
\mathbf{h}%
\end{array}%
\right) \left( \relax{\rm 1\kern-.35em 1},i\relax{\rm 1\kern-.35em 1}\right)
,
\end{eqnarray}%
which, by (\ref{ress}) implies
\begin{eqnarray}
\mathcal{M}\mathbf{L} &=&\mathcal{M}\frac{1}{\sqrt{2}}(\mathcal{B}+\mathcal{B%
}^{\ast })=\frac{1}{\sqrt{2}}\left(
\begin{array}{cc}
-i(\mathbf{h}-\mathbf{h}^{\ast }) & \mathbf{h}+\mathbf{h}^{\ast } \\
i(\mathbf{f}-\mathbf{f}^{\ast }) & -(\mathbf{f}+\mathbf{f}^{\ast })%
\end{array}%
\right) =  \notag \\
&=&\left(
\begin{array}{cc}
0 & -\relax{\rm 1\kern-.35em 1} \\
\relax{\rm 1\kern-.35em 1} & 0%
\end{array}%
\right) \mathbf{L}\left(
\begin{array}{cc}
0 & -\relax{\rm 1\kern-.35em 1} \\
\relax{\rm 1\kern-.35em 1} & 0%
\end{array}%
\right) =\Omega \mathbf{L}\Omega \ ~\blacksquare
\end{eqnarray}

By sandwiching (\ref{defM1}) with the dyonic charge vector $Q$, one also
obtains
\begin{equation}
V_{BH}=-\frac{1}{2}Q^{t}\mathcal{M}(\mathcal{N})Q=\frac12\,(\mathbf{L}%
^{-1}Q)^{T}(\mathbf{L}^{-1}Q)=\frac12\,\,Z^{T}\cdot Z
\end{equation}%
where the real central charge vector $Z$  satisfies
\begin{equation}
Z=\mathbf{L}^{-1}Q\,,  \label{ucentral}
\end{equation}%
with the electric and magnetic real components of $%
Z=(Z_{(m)}^{0},Z_{(m)}^{a},Z_{0}^{(e)},Z_{a}^{(e)})^{T}$ given by universal
formulae in terms of 5D axion and dilation fields
\begin{eqnarray}
Z_{0}^{(e)} &=&e^{-3\phi }(q_{0}+q_{I}a^{I}+\frac{d}{2}p^{0}-\frac{1}{2}%
p^{I}d_{I})\ ,\nn  \label{realem} \\
Z_{I}^{(e)} &=&e^{-\phi }(q_{I}+\frac{1}{2}p^{0}d_{I}-p^{J}d_{IJ})\ ,\nn \\
Z_{(m)}^{0} &=&e^{3\phi }p^{0}\ ,\nn \\
Z_{(m)}^{I} &=&e^{\phi }(p^{I}-p^{0}a^{I})\ ,  \label{cariche}
\end{eqnarray}%
which were derived in \cite{Ceresole:2009id} for $N=8$, but that we can here
interpret as valid for all generalized $d$-geometries. The components with
flat indices are obtained by
\begin{equation}
Z_{a}^{(e)}=Z_{I}^{(e)}(a^{-1/2})_{\phantom{I}a}^{I}\,\qquad ,\qquad
Z_{(m)}^{a}=Z_{(m)}^{I}(a^{1/2})_{I}^{a}\,  \label{flatreal}
\end{equation}%
so that the complex central charge vector with flat indices is
\begin{equation}
Z_{A}=\left(
\begin{array}{c}
Z_{0} \\
Z_{a}%
\end{array}%
\right) =\frac{1}{\sqrt{2}}\left(
\begin{array}{c}
Z_{0}^{(e)}+iZ_{(m)}^{0} \\
Z_{a}^{(e)}+iZ_{(m)}^{a}\label{flatones}%
\end{array}%
\right)
\end{equation}%
and the effective black hole potential is written as \cite{Ceresole:2009id}
\begin{equation}
V_{BH}=|Z_{0}|^{2}+Z_{a}\overline{Z}_{a}\,.
\end{equation}%
%
%
%
%
%
%

\section{\label{Ids-5D}$5D$-Covariant Identities}

In the 5D covariant formalism introduced in \cite{Ceresole:2009id}, it was
found that the kinetic vector matrix $\mathcal{N}_{\Lambda\Sigma}$ in $N=8$,
$D=4$ supergravity can be decomposed as:
\begin{equation}
\mathrm{Re}\mathcal{N}=\left(
\begin{array}{cc}
\frac{d}{3} & -\frac{d_{I}}{2} \\
-\frac{d_{J}}{2} & d_{IJ}%
\end{array}%
\right) \ ,\quad \mathrm{Im}\mathcal{N}=\left(
\begin{array}{cc}
-e^{6\phi }-e^{2\phi }a^{I}a^{J}a_{IJ}\ \ \  & a_{IJ}a^{J} \\
a_{IJ}a^{I} & -e^{2\phi }a_{IJ}%
\end{array}%
\right) \ .
\end{equation}

In virtue of the discussion of Sec. \ref{sectADG}, these formulae hold for
any d-geometry. Note that $\mathrm{Im}\mathcal{N}$ depends on the axions $%
a^{I}$ but not on $d_{IJK}$, whereas $\mathrm{Re}\mathcal{N}$ only depends
on axions, and only through $d_{IJK}$. It is immediate to realize that this
is a consequence of the solvable decomposition (\ref{partial-Iwa}) of $%
\mathbf{L}$, as well as of the relation (\ref{defM1}) between $\mathcal{M}$
and $\mathbf{L}$. Indeed, using (\ref{R}), the matrix $\mathcal{A}$ (\ref{A}%
) can be rewritten as
\begin{equation}
\mathcal{A}=\left(
\begin{array}{cc}
\relax{\rm 1\kern-.35em 1} & 0 \\
\mathrm{Re}\mathcal{N} & \relax{\rm 1\kern-.35em 1}%
\end{array}%
\right) \left(
\begin{array}{cc|cc}
1 & 0 & 0 & 0 \\
a^{I} & 1 & 0 & 0 \\ \hline
0 & 0 & 1 & -a^{J} \\
0 & 0 & 0 & 1%
\end{array}%
\right) \equiv (\mathcal{R})^{-1}\mathcal{A}_{D}(a^{I})\ ,  \label{AD}
\end{equation}%
thus yielding
\begin{equation}
\mathbf{L}=(\mathcal{R})^{-1}\mathcal{A}_{D}\mathcal{D}G\ .
\end{equation}%
Then, since $\mathcal{D}G$ is a diagonal matrix, (\ref{defM1}) implies
\begin{equation}
\mathcal{M}=-(\mathbf{L}^{T})^{-1}\mathbf{L}^{-1}=-(\mathcal{R})^{T}\left[ (%
\mathcal{A}_{D}^{T})^{-1}(\mathcal{D}G)^{-1}(\mathcal{D}G)^{-1}\mathcal{A}%
_{D}^{-1}\right] \mathcal{R}\ .
\end{equation}%
Using (\ref{D}), (\ref{G}) and (\ref{AD}), one can check that
\begin{equation}
-(\mathcal{A}_{D}^{T})^{-1}(\mathcal{D}G)^{-1}(\mathcal{D}G)^{-1}\mathcal{A}%
_{D}^{-1}=\left(
\begin{array}{cc}
\mathrm{Im}\mathcal{N} & 0 \\
0 & \mathrm{Im}\mathcal{N}^{-1}%
\end{array}%
\right) \ .
\end{equation}%
As mentioned, this explains the dependence of $\mathrm{Im}\mathcal{N}$ on
axions alone and not on the $d$-tensor, and that of $\mathrm{Re}\mathcal{N}$
on axions only through $d_{IJK}$.

\section{\label{N=4-D=4}A related case : $N=4$, $D=4$ pure Supergravity}

Although pure 4D $N=4$ supergravity cannot be obtained from five dimensions
by Kaluza-Klein reduction, which would always give rise to the coupling to
matter multiplets, 
we mention it here because of the recent related work of \cite%
{Ferrara:2012ui} and as a simple instance of the splitting of scalar fields
associated with (\ref{partial-Iwa}). The vector kinetic matrix $\mathcal{N}%
_{\Lambda \Sigma }$ in this case reads \cite{CSF} ($\Lambda ,\Sigma =1,...,6$%
)
\begin{equation}
\mathcal{N}_{\Lambda \Sigma }=-S\delta _{\Lambda \Sigma },  \label{N=4}
\end{equation}%
where the axio-dilatonic complex scalar field $S$ of the gravity multiplet,
spanning the rank-$1$ symmetric coset $G/H=SL(2,\mathbb{R})/SO(2)$, is
defined as%
\begin{equation}
S\equiv ie^{\phi }+a\ ,
\end{equation}%
yielding
\begin{equation}
\mathrm{Re}\mathcal{N}_{\Lambda \Sigma }=-a\delta _{\Lambda \Sigma }\
,\qquad \mathrm{Im}\mathcal{N}=-e^{\phi }\delta _{\Lambda \Sigma }\ .
\end{equation}

A solvable basis can be defined also for this theory as in (\ref{N=4}), and
it is given by the \textit{axio-dilatonic symplectic frame } , where the
relevant matrices read
\begin{eqnarray}
\mathcal{M} &=&\left(
\begin{array}{cc}
-e^{\phi }-a^{2}e^{-\phi }\ \  & -ae^{-\phi } \\
-a\,e^{-\phi } & -e^{-\phi }%
\end{array}%
\right) \ ; \\
\mathbf{L} &=&\left(
\begin{array}{cc}
1 & 0 \\
-a & 0%
\end{array}%
\right) \left(
\begin{array}{cc}
e^{-\phi /2} & 0 \\
0 & e^{\phi /2}%
\end{array}%
\right) =\left(
\begin{array}{cc}
e^{-\phi /2} & 0 \\
-a\,e^{-\phi /2} & e^{\phi /2}%
\end{array}%
\right) ,  \label{L}
\end{eqnarray}%
such that the coset representative $\mathbf{L}$ of $SL(2,\mathbb{R})/SO(2)$
satisfies
\begin{equation}
\mathbf{L}^{-1}(a,\phi )=\mathbf{L}(-a,-\phi )\ .
\end{equation}%
In this case the axionic generator%
\begin{equation}
\mathbf{A\equiv }\frac{\partial }{\partial a}\left(
\begin{array}{cc}
1 & 0 \\
-a & 0%
\end{array}%
\right) =\left(
\begin{array}{cc}
0 & 0 \\
-1 & 0%
\end{array}%
\right)
\end{equation}%
is nilpotent of order two rather than of order four, as for generic
d-geometries :%
\begin{equation}
\mathbf{A}^{2}=0.
\end{equation}

The different degree of nilpotency is due to the fact that this theory does
not admit a 5D uplift and thus it is not a $d$-geometry in absence of matter
coupling.

\section{\label{Symmetric}\textit{Vielbein} and $H$-Connection in the Axion
Basis}

When the d-geometry is not only an homogeneous but a \textit{symmetric}
cosets $G/H$ , the \textit{Vielbein} $P_{\mu }$ and H-connection $\omega
_{\mu }$ in a solvable decomposition can be simply computed from the $\left(
\mathfrak{g}\ominus \mathfrak{h}\right) $-valued Maurer-Cartan $1$-form $%
\mathbf{L}^{-1}d\mathbf{L}$ by standard methods
\begin{eqnarray}
(\mathbf{L}^{-1}d\mathbf{L})_{s} &=&\frac{1}{2}\left( \mathbf{L}^{-1}d%
\mathbf{L}+(\mathbf{L}^{-1}d\mathbf{L})^{T}\right) =P_{\mu }\ ;  \label{P} \\
(\mathbf{L}^{-1}d\mathbf{L})_{a} &=&\frac{1}{2}\left( \mathbf{L}^{-1}d%
\mathbf{L}-(\mathbf{L}^{-1}d\mathbf{L})^{T}\right) =\omega _{\mu }\ ,
\label{omega}
\end{eqnarray}%
where subscripts \textquotedblleft $s$" and \textquotedblleft $a$" denote
the symmetric and antisymmetric part, respectively.

The simplest example is provided by the axio-dilatonic coset $G/H=SL(2,%
\mathbb{R})/SO(2)$ treated above, whose coset representative is given by (%
\ref{L}), with Maurer-Cartan $1$-form%
\begin{equation}
\mathbf{L}^{-1}d\mathbf{L}=\left(
\begin{array}{cc}
-\frac{1}{2}d\phi  & 0 \\
-e^{-\phi }da & \frac{1}{2}d\phi
\end{array}%
\right) \ ,
\end{equation}%
leading to the \textit{Vielbein} $P_{\mu }$ and $U(1)$-connection $\omega
_{\mu }$ respectively given by
\begin{equation}
P_{\mu }=\left(
\begin{array}{cc}
-\frac{1}{2}d\phi  & -\frac{1}{2}e^{-\phi }da \\
-\frac{1}{2}e^{-\phi }da & \frac{1}{2}d\phi
\end{array}%
\right) \ ,\qquad \omega _{\mu }=\left(
\begin{array}{cc}
0 & \frac{1}{2}e^{-\phi }da \\
-\frac{1}{2}e^{-\phi }da & 0%
\end{array}%
\right) \ .
\end{equation}%
In particular, one sees that the $U(1)$ connection $\omega _{\mu }$ contains
only the $da$ differential. The kinetic term for the nonlinear $\sigma $%
-model $SL(2,\mathbb{R})/SO(2)$ therefore reads \cite{CSF}
\begin{equation}
\text{Tr}\left( P^{T}P\right) =\frac{1}{2}\left( d\phi ^{2}+e^{-2\phi
}da^{2}\right) \ .
\end{equation}


We now consider in particular $N=8$ supergravity, where the Cartan
decomposition for the $D=4$ scalar manifold (\ref{D=4-coset}) reads%
\begin{eqnarray}
\mathfrak{g} &=&\mathfrak{h}\oplus \mathfrak{k}; \\
\mathfrak{g} &=&\mathfrak{e}_{7(7)};~\mathfrak{h}=\mathfrak{su}(8);~%
\mathfrak{k}=\mathbf{70}\text{ of~}\mathfrak{su}(8)\text{.}
\end{eqnarray}%
According to (\ref{dec-1})-(\ref{dec-2}), the following $\mathfrak{usp}(8)$%
-covariant branchings take place:
\begin{eqnarray}
\mathfrak{k} &:&\mathbf{70}=\mathbf{1}_{\mathfrak{k}}+\mathbf{42}_{\mathfrak{%
k}}+\mathbf{27}_{\mathfrak{k}}; \\
\mathfrak{h} &:&\mathbf{63}_{\mathfrak{h}}=\mathbf{36}_{\mathfrak{h}}+%
\mathbf{27}_{\mathfrak{h}}
\end{eqnarray}%
The coset \textit{Vielbein} $P_{\mu }$ is given by the non-compact
generators
\begin{eqnarray}
\mathbf{1}_{\mathfrak{k}} &:&\mathcal{D}^{-1}\partial \mathcal{D};  \notag \\
\mathbf{42}_{\mathfrak{k}} &:&[\mathcal{G}^{-1}\partial\mathcal{G}]_{s};
\notag \\
\mathbf{27}_{\mathfrak{k}} &:&\left[ (\mathcal{D}\mathcal{G})^{-1}\partial
T(a)(\mathcal{D}\mathcal{G})\right] _{s},
\end{eqnarray}%
while the compact ones give the $SU(8)$-connection $\omega _{\mu }$
\begin{eqnarray}
\mathbf{36}_{\mathfrak{h}} &\rightarrow &[\mathcal{G}^{-1}\partial \mathcal{G%
}]_{a};  \notag \\
\mathbf{27}_{\mathfrak{h}} &\rightarrow &\left[ (\mathcal{D}\mathcal{G}%
)^{-1}\partial T(a)(\mathcal{D}\mathcal{G})\right] _{a}.
\end{eqnarray}

The Maurer-Cartan $1$-form gets generally decomposed as
\begin{equation}
\mathbf{L}^{-1}\partial \mathbf{L}=(\mathcal{D}\mathcal{G})^{-1}\partial
T(a)(\mathcal{D}\mathcal{G})+\mathcal{D}^{-1}\partial \mathcal{D}+\mathcal{G}%
^{-1}\partial\mathcal{G}\ .  \label{L-bold}
\end{equation}%
From the definitions (\ref{A}), (\ref{D}) and (\ref{G}), one can compute
\begin{eqnarray}
\mathcal{D}^{-1}\partial \mathcal{D} &=&\left(
\begin{array}{cc|cc}
-3 & 0 & 0 & 0 \\
0 & -1 & 0 & 0 \\ \hline
0 & 0 & 3 & 0 \\
0 & 0 & 0 & 1%
\end{array}%
\right) d\phi \ =\left( \mathcal{D}^{-1}\partial \mathcal{D}\right) _{s}; \\
\mathcal{G}^{-1}\partial \mathcal{G }&=&\left(
\begin{array}{cc|cc}
0 & 0 & 0 & 0 \\
0 & E^{-1}dE & 0 & 0 \\ \hline
0 & 0 & 0 & 0 \\
0 & 0 & 0 & -E^{-1}dE%
\end{array}%
\right) \ ; \\
(\mathcal{D}\mathcal{G})^{-1}\partial T(a)(\mathcal{D}\mathcal{G})
&=&e^{-2\phi }\left(
\begin{array}{cc|cc}
0 & 0 & 0 & 0 \\
(a^{1/2})_{I}^{a}da^{I} & 0 & 0 & 0 \\ \hline
0 & 0 & 0 & -(a^{1/2})_{I}^{b}da^{I} \\
0 & d_{IJK}(a^{-1/2})_{a}^{J}(a^{-1/2})_{b}^{K}da^{I} & 0 & 0%
\end{array}%
\right) \ .
\end{eqnarray}%
This implies that the Maurer-Cartan 1-form $\mathbf{L}^{-1}\partial \mathbf{L%
}$ does not depend on the axions $a^{I}$ explicitly, but only on their
differential $da^{I}$. 

According to (\ref{P}) and (\ref{omega}), the \textit{Vielbein} $P_{\mu }$
and $SU(8) $-connection $\omega _{\mu }$ for the coset (\ref{D=4-coset}) are
the symmetric and anti-symmetric part of (\ref{L-bold}), respectively. In
particular, the component $\mathbf{27}_{\mathfrak{k}}$ of $P_{\mu }$ and the
component $\mathbf{27}_{\mathfrak{h}}$ of $\omega _{\mu }$ respectively
read:
\begin{eqnarray}
\mathbf{27}_{\mathfrak{k}} &:&\left[ (\mathcal{D}\mathcal{G})^{-1}\partial
T(a)(\mathcal{D}\mathcal{G})\right] _{s}=  \notag \\
&=&\frac{1}{2}e^{-2\phi }\left(
\begin{array}{cc|cc}
0 & (a^{1/2})_{I}^{b}da^{I} & 0 & 0 \\
(a^{1/2})_{I}^{a}da^{I} & 0 & 0 &
d_{IJK}(a^{-1/2})_{a}^{J}(a^{-1/2})_{b}^{K}da^{I} \\ \hline
0 & 0 & 0 & -(a^{1/2})_{I}^{b}da^{I} \\
0 & d_{IJK}(a^{-1/2})_{a}^{J}(a^{-1/2})_{b}^{K}da^{I} &
-(a^{1/2})_{I}^{a}da^{I} & 0%
\end{array}%
\right) \ ;  \notag \\
&& \\
\mathbf{27}_{\mathfrak{h}} &:&\left[ (\mathcal{D}\mathcal{G})^{-1}\partial
T(a)(\mathcal{D}\mathcal{G})\right] _{a}=  \notag \\
&=&\frac{1}{2}e^{-2\phi }\left(
\begin{array}{cc|cc}
0 & -(a^{1/2})_{I}^{b}da^{I} & 0 & 0 \\
(a^{1/2})_{I}^{a}da^{I} & 0 & 0 &
-d_{IJK}(a^{-1/2})_{a}^{J}(a^{-1/2})_{b}^{K}da^{I} \\ \hline
0 & 0 & 0 & -(a^{1/2})_{I}^{b}da^{I} \\
0 & d_{IJK}(a^{-1/2})_{a}^{J}(a^{-1/2})_{b}^{K}da^{I} &
(a^{1/2})_{I}^{a}da^{I} & 0%
\end{array}%
\right) \ .  \notag \\
&&
\end{eqnarray}

\section{\label{Flat}Flat Connections and Axion Basis}

As shown in \cite{Strominger-SKG} and further investigated in \cite{Lerche-1}%
, the defining identities of $N=2$ special K\"{a}hler geometry can be viewed
as the flatness condition of a non-holomorphic connection $\mathcal{A}_{I}$
and can be encoded into a first-order matrix equation \cite{Lerche-1}%
\begin{equation}
\left( \partial _{i}-\mathcal{A}_{i}\right) \mathbf{U}=0\, ,  \label{Strom-1}
\end{equation}%
where $\mathbf{U}$ is a non-holomorphic matrix $(V, D_iV, \overline
D_{\bar\imath}\overline V, \overline V ) $ with $V=(X^\Lambda,F_\Lambda)$.
One can further choose a gauge where $\mathcal{A}_{i}$ becomes holomorphic%
\begin{equation}
\overline{\mathcal{A}}_{i}=0\Rightarrow \mathcal{A}_{i}=\mathtt{A}_{i}\text{%
,~}\overline{\partial }\mathtt{A}_{i}=0,  \label{holo}
\end{equation}%
such that (\ref{Strom-1}) can be recast as follows:%
\begin{equation}
\left( \partial _{i}-\mathtt{A}_{i}\right) \mathbf{V}=0,  \label{Strom-2}
\end{equation}%
with now an holomorphic solution matrix $\mathbf{V}$ containing $V$ in the
first row. In turn, the holomorphic flat connection $\mathtt{A}_{i}$ can be
decomposed as%
\begin{equation}
\mathtt{A}_{i}=\mathbf{\Gamma }_{i}+\mathbf{C}_{i},
\end{equation}%
where $\mathbf{\Gamma }_{i}$ is the diagonal part (which vanishes in \textit{%
special coordinates}), and $\mathbf{C}_{i}$ generates an Abelian subalgebra
of $\mathfrak{sp}(2n+2,\mathbb{R})$ that is nilpotent of order four:%
\begin{equation}
\mathbf{C}_{i}\mathbf{C}_{j}\mathbf{C}_{k}\mathbf{C}_{l}=0.
\end{equation}

The case of special K\"{a}hler $d$-geometry in the axion basis basis is
analysed in App. C of \cite{Lerche-1}. In particular, by recalling (\ref{T}%
), one can compute the axionic generators of the solvable parametrization of
the $D=4$ scalar manifold treated above as
\begin{equation}
\partial T(a)/\partial a^{k}=\left(
\begin{array}{cc|cc}
0 & 0 & \ 0 & 0 \\
\delta _{k}^{j} & 0 & \ 0 & 0 \\ \hline
0 & 0 & \ 0 & -\delta _{k}^{i} \\
0 & d_{ijk} & \ 0 & 0%
\end{array}%
\right) .  \label{dT}
\end{equation}%
Up to relabelling of rows and columns, (\ref{dT}) matches the expression of $%
\mathbf{C}_{i}$ (for $n=27$) given by (3.6) of \cite{Lerche-1}.

For $N=2$ special K\"{a}hler $d$-geometries (namely, for those special
geometries admitting an uplift to $D=5$) in the axion basis, this highlights
the relation between the solvable parametrization of the $D=4$ scalar
manifold discussed in Sec. \ref{sectADG} and the nilpotent connection of the
reformulation \textit{\`{a} la Strominger} in the holomorphic gauge (\ref%
{holo}).


\section{\label{N=2}$N=2$ Special K\"{a}hler $d$-Geometry, Symplectic
Sections and the Unitary Matrix $M$}


In this section we are going to make contact with $N=2$ special K\"{a}hler $%
d $-geometries \cite{dWVVP} in the symplectic frame defined by the cubic
prepotential (\ref{FF}). We recall for convenience some results of \cite%
{CFM1} and we build on them. 
It has already been remarked that $N=2$ special K\"{a}hler $d$-geometry
differs from the higher $N$-extended theories in that the $n_{V}$ 5D axions $%
a^{i}$
exactly combine with the 5D scalars $\lambda ^{i}=\lambda ^{i}(\lambda
^{x},\phi )$ in order to give complex 4D scalar fields $\frac{X^{i}}{X^{0}}%
=z^{i}\equiv a^{i}-i\lambda ^{i}$, where $X^{\Lambda }=X^{0},X^{i}$.
Moreover, in ${N}=2$ the central charge can be readily computed from the
cubic prepotential $F(X)$ of eq. (\ref{FF}) by the usual formula (\ref%
{centralcharge})
\begin{equation}
Z=e^{\frac{K(z,{\bar{z}})}{2}}(X^{\Lambda }q_{\Lambda }-F_{\Lambda
}p^{\Lambda })\,
\end{equation}%
%
For $N=2$ cubic geometry one finds\cite{CFM1}
\begin{eqnarray}
Z &=&\frac{1}{\sqrt{8\mathcal{V}}}[q_{0}+q_{i}z^{i}+p^{0}f(z)-p^{i}f_{i}(z)]%
\ ;  \label{Z2} \\
D_{i}Z &=&(\partial _{i}+\frac{1}{2}\partial _{i}K)Z=\frac{1}{\sqrt{8%
\mathcal{V}}}\left[ q_{0}\partial _{i}K+q_{j}(\delta _{i}^{j}+\partial
_{i}K\,z^{j})\right. +\nn  \notag \\
&&\left. +p^{0}\left( f_{i}(z)+\partial _{i}K\,f(z)\right)
-p^{j}(f_{ij}(z)+\partial _{i}K\,f_{j}(z))\right] \ ,  \label{DaZ2}
\end{eqnarray}%
where
\begin{equation}
f(z)=\frac{1}{3!}d_{ijk}z^{i}z^{j}z^{k}\ ,\quad f_{i}(z)=\frac{1}{2}%
d_{ijk}z^{j}z^{k}\ ,\quad f_{ij}(z)=d_{ijk}z^{k}\,,\quad \mathcal{V}=\frac{1%
}{3!}d_{ijk}\lambda ^{i}\lambda ^{j}\lambda ^{k}=e^{6\phi }\ ,
\end{equation}%
with the (real) K\"{a}hler potential and its (purely imaginary) derivatives
given by
\begin{equation}
K=-\ln (8\mathcal{V})\ ;\qquad \qquad \partial _{i}K=-\frac{i}{4\mathcal{V}}%
d_{ijk}\lambda ^{j}\lambda ^{k}=-\partial _{\bar{\imath}}K\ .
\end{equation}%
Notice that $i$ is a \textit{curved} index of the 5D U-duality group $G_{5}$%
, and $\Lambda =(0,i)$. The connection with the universal basis is given by
introducing $n_{V}$ 5D scalars as $\hat{\lambda}^{i}=e^{-2\phi }\lambda ^{i}$
so that they satisfy $d_{ijk}{\hat{\lambda}}^{i}{\hat{\lambda}}^{j}{\hat{%
\lambda}}^{k}=1$. The $n_{V}$ complex 4D scalar components are then $%
(a^{i},\phi ,\hat{\lambda}^{i})$ . The special K\"{a}hler metric is given by
\begin{eqnarray}
\qquad g_{ij} &=&\frac{1}{4}(\frac{1}{4}{\kappa }_{i}{\kappa }_{j}-{\kappa }%
_{ij})\mathcal{V}^{-2/3}=\frac{1}{4}\mathcal{V}^{-2/3}a_{ij}=\frac{1}{4}%
e^{-4\phi }a_{ij}\ , \\
{\kappa }_{i} &=&\mathcal{V}^{-2/3}d_{ijk}\lambda ^{j}\lambda ^{k}\ ,\qquad {%
\kappa }_{ij}=\mathcal{V}^{-1/3}d_{ijk}\lambda ^{k}\ .
\end{eqnarray}%
\noindent One can assemble $Z$ and $\overline{D}_{\bar{\imath}}\overline{Z}$
into a symplectic central charge vector $Z_{\alpha }$ with a curved lower
index
\begin{eqnarray}
{\mathcal{Z}}_{\alpha } &=&\left(
\begin{array}{c}
Z \\
\overline{D}_{\bar{\imath}}\overline{Z}%
\end{array}%
\right) \equiv \langle Q,V_{\alpha }\rangle =Q^{T}\Omega V_{\alpha }=f_{%
\phantom\Lambda \alpha }^{\Lambda }q_{\Lambda }-h_{\Lambda \,\alpha
}p^{\Lambda },  \label{c-1} \\
V_{\alpha } &=&\left(
\begin{array}{c}
f_{\phantom\Lambda \alpha }^{\Lambda } \\
h_{\Lambda \alpha }%
\end{array}%
\right) \,.  \label{Q-V}
\end{eqnarray}%
%
%
%
%
%
%
%
%
%
%
%
%
%
%
%
%
%
%
%
%
%
\noindent Then, from ${\mathcal{Z}}_{\alpha }$ in (\ref{Z2}) and (\ref{DaZ2}%
) one can read off the components of $V_{\alpha }$, which are
\begin{eqnarray}
\mathbf{f}\equiv f_{\phantom\Lambda \alpha }^{\Lambda } &=&({f}_{\phantom%
\Lambda 0}^{\Lambda }\,,~f_{\phantom\Lambda {\bar{\jmath}}}^{\Lambda })=%
\frac{1}{\sqrt{8\mathcal{V}}}\left(
\begin{array}{cc}
1 & \partial _{\bar{\jmath}}K \\
z^{i}\ \  & \delta _{\bar{\jmath}}^{\bar{\imath}}+\partial _{\bar{\jmath}}K\,%
\bar{z}^{\bar{\imath}}%
\end{array}%
\right) \ ;  \label{f-hat-N=2} \\
\mathbf{h}\equiv h_{\Lambda \alpha } &=&({h}_{\Lambda 0}\,,~{h}_{\Lambda {%
\bar{\jmath}}})=\frac{1}{\sqrt{8\mathcal{V}}}\left(
\begin{array}{cc}
-f(z)\ \  & -\overline{f}_{\bar{\jmath}}(\bar{z})-\partial _{\bar{\jmath}}K%
\overline{f}(\bar{z}) \\
f_{i}(z)\ \  & \bar{f}_{\bar{\imath}\,\bar{\jmath}}(\bar{z})+\partial _{\bar{%
\jmath}}K\,\overline{f}_{\bar{\imath}}(\bar{z})%
\end{array}%
\right) .
\end{eqnarray}

While it can be checked that%
\begin{equation}
i(\mathbf{f^{\dag }h-h^{\dag }f})_{{\alpha \beta }}={\mathcal{G}}_{\alpha
\beta }=\left(
\begin{array}{cc}
1 & 0 \\
0 & g_{ij}%
\end{array}%
\right) ,
\end{equation}%
we should better consider the normalized symplectic sections with flat
tangent indices ${A}=(0,{a})$, such that
\begin{equation}
i(\mathbf{f^{\dag }h-h^{\dag }f})_{{A}{B}}=\delta _{{A}{B}}.
\end{equation}%
They are the components of ${\mathcal{Z}}_{A}=(Z,\overline{D}_{\overline{a}}%
\overline{Z})$, and they can be obtained by flattening the curved indices $i$
by the $G_{5}$- \textit{Vielbein} $e_{i}^{{a}}$$\,$\footnote{%
Further below, in the explicit case of $stu$ model, the \textit{Vielbein}
will be taken to be purely imaginary (\textit{cfr.} App. \ref{App-1}).}%
so that the orthonormalized symplectic sections $f_{\phantom\Lambda {A}%
}^{\Lambda }$ and $h_{\Lambda {A}}$ are given by
\begin{equation}
f_{\phantom\Lambda A}^{\Lambda }=f_{\phantom\Lambda \,\alpha }^{\Lambda
}(G^{-1/2})_{\phantom\alpha A}^{\alpha }\ ,\qquad h_{\Lambda A}=h_{\Lambda
\,\alpha }(G^{-1/2})_{\phantom\alpha A}^{\alpha }\,.  \label{rel-N=2}
\end{equation}%
%
%
%
%
%
%
%
%
%
%
%
%
%
%
%
%
%
%
%
%
%
%
%
%
%
%
%
%
%
%
%
%

It was emphasized in \cite{Ceresole:2009id} that the symplectic sections $%
\mathbf{f}$\textbf{\ }and $\mathbf{h}$ of (generalized) special geometry are
defined only up to the action
\begin{equation}
\mathbf{f}\rightarrow \mathbf{f}^{\prime }\equiv \mathbf{f}M\ ,\qquad
\mathbf{h}\rightarrow \mathbf{h}^{\prime }\equiv \mathbf{h}%
M~~\Leftrightarrow ~M=\mathbf{f}^{-1}\mathbf{f}^{\prime }=\mathbf{h}^{-1}%
\mathbf{h}^{\prime },  \label{M-action}
\end{equation}%
of a \textit{unitary} matrix $M$, which preserves the form of the kinetic
vector matrix ${\mathcal{N}}={\mathbf{h}}{\mathbf{f}}^{-1}$ and the
conditions (\ref{prop}) derived from symplectic invariance of $\mathbf{L}$.
Actually, the matrix $M$ found in \cite{Ceresole:2009id} to connect $N=2$
with $N=8$ is exactly the necessary one to rotate the usual basis of special
geometry into the axion basis of any $d$-geometry. It can be written as
\begin{eqnarray}
M &=&\frac{1}{2}\left(
\begin{array}{cc}
1 & (g^{-1/2})_{\phantom{j}\tilde{a}}^{\,\bar{\jmath}}\partial _{\bar{\jmath}%
}K \\
-i\mathcal{V}^{-1/3}\lambda ^{i}(a^{1/2})_{i}^{\phantom ia}\ \ \  & \left(
\mathcal{V}^{-1/3}\delta _{\bar{\jmath}}^{i}+i\mathcal{V}^{-1/3}\lambda
^{i}\partial _{\bar{\jmath}}K\right) (a^{1/2})_{i}^{\phantom ia}(g^{-1/2})_{%
\phantom j\tilde{a}}^{j}%
\end{array}%
\right) \ ;  \label{MM} \\
MM^{\dag } &=&\relax{\rm 1\kern-.35em 1},  \label{MM-2}
\end{eqnarray}%
where%
\begin{eqnarray}
\partial _{\bar{\jmath}}K &=&2i\lambda ^{i}g_{i\bar{\jmath}}\ ; \\
g_{ij} &=&\frac{1}{4}\mathcal{V}^{-2/3}a_{ij}\ ; \\
(g^{-1/2})_{a}^{\bar{\jmath}}\partial _{\bar{\jmath}}K &=&2i\lambda
^{i}(g^{1/2})_{i}^{b}\delta _{ab}\ ; \\
(g^{-1/2})_{a}^{i} &=&2\mathcal{V}^{1/3}(a^{-1/2})_{a}^{i}\ .
\end{eqnarray}


By further rescaling the $D=4$ dilatons as%
\begin{equation}
\lambda ^{i}\equiv \mathcal{V}^{1/3}\hat{\lambda}^{i},~\frac{1}{6!}d_{ijk}%
\hat{\lambda}^{i}\hat{\lambda}^{j}\hat{\lambda}^{k}=1.
\end{equation}%
the matrix $M$ (\ref{MM}) can be recast as follows:
\begin{equation}
M=\frac{1}{2}\left(
\begin{array}{cc}
1 & i\hat{\lambda}^{i}(a^{1/2})_{i}^{b}\delta _{ab} \\
-i\hat{\lambda}^{i}(a^{1/2})_{i}^{\phantom ia}\ \ \  & 2\delta _{\tilde{a}%
}^{a}-\hat{\lambda}^{i}\hat{\lambda}^{j}(a^{1/2})_{i}^{\phantom %
ia}(a^{1/2})_{j}^{\phantom ib}\delta _{\tilde{a}b}%
\end{array}%
\right) \ .  \label{MM-recast}
\end{equation}%
Using (\ref{deff}), one can see that the action (\ref{M-action}) of $M$
induces the following transformation of the coset representative $\mathbf{L}$%
:
\begin{equation}
\mathbf{L}\rightarrow \mathbf{L}^\prime {}=\mathbf{L}\left(
\begin{array}{cc}
\text{Re}M & -\text{Im}M \\
\text{Im}M & \text{Re}M%
\end{array}%
\right) \ \equiv \mathbf{L}\mathcal{Y}\left( \text{Re}M,\text{Im}M\right) ,
\label{Y-call}
\end{equation}%
where the real symmetric and unitary matrix
\begin{eqnarray}
\mathcal{Y} &=&\frac{1}{2}\left(
\begin{array}{cc|cc}
1 & 0 & 0 & -\hat{\lambda}^{i}(a^{1/2})_{i}^{b}\delta _{ab} \\
0 & 2\delta _{b}^{a}-\hat{\lambda}^{i}\hat{\lambda}%
^{j}(a^{1/2})_{i}^{a}(a^{1/2})_{j}^{c}\delta _{bc} & \hat{\lambda}%
^{i}(a^{1/2})_{i}^{a} & 0 \\ \hline
0 & \hat{\lambda}^{i}(a^{1/2})_{i}^{b}\delta _{ab} & 1 & 0 \\
-\hat{\lambda}^{i}(a^{1/2})_{i}^{a} & 0 & 0 & 2\delta _{b}^{a}-\hat{\lambda}%
^{i}\hat{\lambda}^{j}(a^{1/2})_{i}^{a}(a^{1/2})_{j}^{c}\delta _{bc}%
\end{array}%
\right) \ ;  \notag \\
&&  \label{YY} \\
&&\mathcal{Y}^{\ast }=\mathcal{Y}^{\dag }=\mathcal{Y}^{T}=\mathcal{Y}%
^{-1}\Leftrightarrow \mathcal{YY}^{\dag }=\mathcal{YY}^{T}=\mathcal{Y}^{2}=%
\relax{\rm 1\kern-.35em 1},  \label{YY-1}
\end{eqnarray}%
does not depend on the volume modulus $\mathcal{V}$.

The symplecticity of $\mathbf{L}$ (and thus of $\mathbf{L}^\prime $) yields
\begin{equation}
\mathbf{L}^\prime {}^{T}\Omega \mathbf{L}^\prime {}=\Omega \rightarrow
\mathcal{Y}^{T}\Omega \mathcal{Y}=\Omega \ ,  \label{symplcondY}
\end{equation}%
thus also $\mathcal{Y}$ is a symplectic matrix, as expected. Indeed, from
its very definition (\ref{Y-call}), the symplectic condition (\ref%
{symplcondY}) becomes
\begin{equation}
\mathrm{Im}M\,\mathrm{Re}M+\mathrm{Re}M\,\mathrm{Im}M=0\ ,\qquad \mathrm{Re}%
M^{2}-\mathrm{Im}M^{2}=\relax{\rm 1\kern-.35em 1}\ ,  \label{UnitConstr}
\end{equation}%
which is identically satisfied since $M$ is a unitary matrix, with $\mathrm{%
Re}M^{T}=\mathrm{Re}M$\ ,\ \ and \ \ $\mathrm{Im}M^{T}=~-\mathrm{Im}M$ (%
\textit{cfr.} (\ref{MM})-(\ref{MM-2})).\medskip

The matrix $\mathcal{Y}\left( \text{Re}M,\text{Im}M\right) $ (\ref{Y-call})
provides a realization of the maximal symmetric embedding \cite%
{Gaillard:1981rj}%
\begin{equation}
U(28)\subset Sp(56,\mathbb{R}).  \label{emb}
\end{equation}

Indeed, since $\mathbf{L}$ is symplectic, one has checked that also $%
\mathcal{Y}$ is symplectic, but given (\ref{YY-1}), this leads to
\begin{equation}
\left[ \mathcal{Y},\Omega \right] =0\ .
\end{equation}%
%

An explicit computation of the matrices $M$ (\ref{MM-recast}) and $\mathcal{Y%
}$ (\ref{YY}) for the $t^{3}$ limit of the $stu$ model is presented in App. %
\ref{Expl-T^3}. 

\section{\label{Sec-2}Unitarity Relations for $M$ and Induced Relations on $%
\hat{M}$}

The residual freedom in the definition of the symplectic section was found
in \cite{Ceresole:2009id} to imply that the symplectic vector ${\mathcal{Z}}%
_{A}=\left( Z,\overline{D}_{\overline{{a}}}\overline{Z}\right) ^{T}$ of $N=2$
special geometry, with a flat index $A=(0,\bar{a})$, differs by a unitary
transformation from the corresponding central charge vector $%
Z_{A}=(Z_{0},Z_{a})^{T}$ of the $N=8$, $D=4$ theory (\ref{flatones}) in the $%
E_{6(6)}$-covariant symplectic frame (with $a=1,...,27$),
\begin{equation}
Z_{A}={\mathcal{Z}}_{B}M_{{\phantom B}A}^{B}\,.
\end{equation}%
This is obvious from the fact that the $N=2$ sections in (\ref{f-hat-N=2})
are not lower triangular, as required in the axion basis in (\ref{2.21})
where the the symplectic section $\mathbf{f}$ is real. Notice that the $%
E_{6(6)}$ basis is related to the usual de Wit and Nicolai symplectic frame
by a symplectic transformation \cite{Ceresole:2009jc} . However, under a
change of symplectic basis, that is a duality transformation, the kinetic
matrix transforms as $\mathcal{N}_{\Lambda \Sigma }\rightarrow (C+D\mathcal{N%
})(A+B\mathcal{N})^{-1}$, while the unitary transformation $M$ leaves $%
\mathcal{N}_{\Lambda \Sigma }$ invariant.

%

$M$ acts on the normalized sections, with a flat tangent index, as given by (%
\ref{M-action}) (where now prime refers to $N=8$ sections and unprimed
sections are the $N=2$ ones in the axion basis, discussed in Sec. \ref{N=2}%
). On the other hand, one can define a matrix $\widehat{M} $ acting on
(un-normalized) sections with a curved lower index as
\begin{equation}
\widehat{\mathbf{f}}^{\prime }=\widehat{\mathbf{f}}\widehat{M} \ , \qquad
\widehat{\mathbf{h}}^{\prime }=\widehat{\mathbf{h}}\widehat{M} \qquad
\Leftrightarrow \qquad \widehat{M}=\widehat{\mathbf{f}}^{-1}\widehat{\mathbf{%
f}} ^{\prime }=\widehat{\mathbf{h}}^{-1}\widehat{\mathbf{h}} ^{\prime } \ ,
\label{M-hat}
\end{equation}

They can be obtained from (\ref{2.21}) and (\ref{2.22}) by multiplication
with the appropriate \textit{Vielbein}, that is
\begin{equation}
\widehat{{f}}_{\phantom\Lambda \alpha }^{\Lambda }={f}_{\phantom\Lambda
A}^{\Lambda }(A^{1/2})_{\alpha }^{A}\ ,\qquad \widehat{{h}}_{\Lambda \alpha
}={h}_{\Lambda \,A}(A^{1/2})_{\alpha }^{A}\,,  \label{rel-N=8}
\end{equation}%
with%
\begin{equation}
A\equiv \left(
\begin{array}{c|c}
1 & 0...0 \\ \hline
&  \\
\vspace{-1pt}%
\begin{array}{c}
0\  \\
...\  \\
0\
\end{array}
& \ \ a_{IJ}\ \
\end{array}%
\right)
\end{equation}%
where $\ a_{IJ}$ is the kinetic vector matrix of $N=8$, $D=5$ supergravity.
In the $E_{6(6)}$-frame of 4D $N=8$ supergravity, the symplectic section
with curved indices $\widehat{\mathbf{f}}$ read \cite{Ceresole:2009id}
\begin{equation}
\hat{f}_{\phantom{\Lambda}\alpha }^{\Lambda }=\frac{1}{\sqrt{2}}\left(
\begin{array}{c|c}
&  \\
e^{-3\phi }\  & \ 0 \\
& \  \\ \hline
&  \\
\vspace{-1pt}e^{-3\phi }a^{J}\  & \ e^{-\phi }\delta _{I}^{J} \\
&  \\
&
\end{array}%
\right) \ ,\quad (\hat{f}^{-1})_{\Lambda }^{\phantom\Lambda \alpha }=\sqrt{2}%
\left(
\begin{array}{c|c}
&  \\
e^{3\phi }\  & \ 0 \\
& \vspace{-1pt} \\ \hline
&  \\
\vspace{-1pt}-e^{\phi }a^{I}\  & \ e^{\phi }\delta _{J}^{I} \\
&  \\
&
\end{array}%
\right) \ ,  \label{f-hat-N=8}
\end{equation}%
where, in the \textit{symmetric} gauge \cite{Ceresole:2009jc}, $\Lambda =0,I$
and $\alpha =0,I$, where here $I$ is a \textit{curved} index spanning the $%
\mathbf{27}$ of $E_{6(6)}$.

From (\ref{f-hat-N=2}), (\ref{f-hat-N=8}) and (\ref{rel-N=8}), one can
compute the matrix \cite{Ceresole:2009id}
\begin{equation}
\hat{M}_{\alpha }^{\beta }=(\hat{f}^{-1})_{\Lambda }^{\phantom\Lambda \beta }%
\hat{f}_{\phantom\Lambda \alpha }^{\prime \Lambda }=\frac{1}{2}\left(
\begin{array}{cc}
1\ \  & \partial _{\bar{\jmath}}K \\
-i\lambda ^{i}\mathcal{V}^{-1/3}\ \  & \mathcal{V}^{-1/3}\delta _{j}^{i}+i%
\mathcal{V}^{-1/3}\lambda ^{i}\partial _{\bar{\jmath}}K%
\end{array}%
\right) ,  \label{MM-hat}
\end{equation}%
which does not depend on the axion fields. Moreover, using (\ref{rel-N=2}), (%
\ref{rel-N=8}) and (\ref{M-action}), the relation between $\hat{M}$ and $M$
is given by%
\begin{equation}
\widehat{M}=\widehat{\mathbf{f}}^{-1}\widehat{\mathbf{f}}^{\prime }=A^{-1/2}%
\mathbf{f}^{-1}\mathbf{f}^{\prime }{\mathcal{G}}^{1/2}=A^{-1/2}M{\mathcal{G}}%
^{1/2}\Leftrightarrow M=A^{1/2}\widehat{M}{\mathcal{G}}^{-1/2}.
\end{equation}

The unitarity of $M$ entails the following identities for $\hat{M}$, namely:
\begin{eqnarray}
MM^{\dag } &=&Id\Leftrightarrow A\,\hat{M}\,{\mathcal{G}}^{-1}\hat{M}^{\dag
}=Id\ ;  \label{id-1} \\
&&  \notag \\
M^{\dag }M &=&Id\Leftrightarrow {\mathcal{G}}^{-1}\,\hat{M}^{\dag }A\,\hat{M}%
=Id\ .  \label{id-2}
\end{eqnarray}

\section{\label{FO}Axion Basis and the Fake Superpotential}

In this section we show an interesting application of the axion basis to
non-BPS extremal black holes. The unitary transformation $M$ that rotates
the usual $N=2$ basis of special geometry ${\mathcal{Z}}_{A}$ into the $%
E_{6(6)}$ basis $Z_{A}$ allows to make a precise connection with the $N=2$ $%
stu$ model, where the three complex scalar fields $z^{i}=\{s\,,t\,,u\}$ span
the rank-$3$ coset space $\left[ \frac{SU(1,1)}{U(1)}\right] ^{3}$, with
\begin{equation}
f=stu\ ,\quad e^{-K}=8\lambda ^{1}\lambda ^{2}\lambda ^{3}=8\mathcal{V}\ ,
\end{equation}%
viewed as a sub sector of the full $N=8$ theory \cite%
{Ferrara:2006em,Ceresole:2009iy,Ceresole:2009vp}. The aim is to illustrate
the computation of the fake superpotential for non-BPS solutions and $%
(p^{0},q_{0})$ charge configuration in the stu-truncation of $N=8$
supergravity. This example was discussed from two different viewpoints: in
\cite{Ceresole:2009iy} the fake superpotential was computed for generic
charges in terms of duality invariants of the underlying special geometry,
while in \cite{Bossard:2009we} Bossard, Michel and Pioline (BMP) provided a
procedure based on nilpotent orbits which lead to the fake superpotential as
solution of a sixth order polynomial.

The virtue of the axion basis is that, while showing the equivalence of the
derivation of \cite{Bossard:2009we} and \cite{Ceresole:2009iy}, we can read
out the fake superpotential from the $N=8$ central charge in the skew
symmetric form. Here we start from the formula for the central charge
derived in \cite{Ceresole:2009id} using 4D/5D special geometry relations,
and we look for a suitable $SU(8)$ transformation that brings it to the form
given by Eq. (2.68) of \cite{Bossard:2009we}
\begin{equation}
Z_{AB}^{{CFG}}\overset{SU(8)}{\longrightarrow }Z_{AB}^{{BMP}}
\end{equation}

In particular, we study the effect of such a rotation with respect to the
decomposition $\mathbf{28}\rightarrow \mathbf{1}_{\mathbb{C}}+\mathbf{27}_{%
\mathbb{C}}$, which is common to the central charge normal frame of both
\cite{Ceresole:2009id} and \cite{Bossard:2009we} . We identify this
transformation in the $t^{3}$-truncation where it depends only on one angle $%
\chi $, purely given in terms of duality invariant quantities. When this
rotation is used to match the central charge in \cite{Ceresole:2009id} and
that of \cite{Bossard:2009we}, we consistently retrieve the non-BPS fake
superpotential for the $N=2$ $t^{3}$ model, within the $(p^{0},q_{0})$
charge configuration in presence of non zero axions. This is a non-trivial
consistency check for the 4D/5D formalism based on the matrices $\widehat{M}$
and $M$ \cite{Ceresole:2009id} detailed in previous Sections.


The key point of this analysis is that the $\mathbf{28}$ components of the $%
N=8$ central charge matrix $Z_{AB}$ can be traded for the symplectic vectors
$Z_{A}$ (with flat lower index) or $Z_{\alpha }$ (with a curved one)
reflecting the splitting $\mathbf{28}=\mathbf{1}_{\mathbb{C}}+\mathbf{27}_{%
\mathbb{C}}$ of the axion basis. Since $Z_{AB}$ can always be brought to the
skew-diagonal form 
\begin{equation}
Z_{AB}=\left(
\begin{array}{cccc}
z_{1} & 0 & 0 & 0 \\
0 & z_{2} & 0 & 0 \\
0 & 0 & z_{3} & 0 \\
0 & 0 & 0 & z_{4}%
\end{array}%
\right) \otimes \epsilon \ ,  \label{skew-diag}
\end{equation}%
one has to relate the eigenvalues $z_{1}$, $z_{2}$, $z_{3}$, $z_{4}$ with
the complex components of $Z_{\alpha }=(Z_{0},Z_{I})$ \cite{Ceresole:2009id}%
, with $I=1,2,3$,
\begin{eqnarray}
Z_{0} &=&\frac{1}{\sqrt{2}}(Z_{0}^{(e)}+iZ_{(m)}^{0})\ ,\nn \\
Z_{I} &=&\frac{1}{\sqrt{2}}(Z_{I}^{(e)}+ia_{IJ}Z_{(m)}^{J})\ .
\end{eqnarray}

In fact, in light of the previous discussion, Eqs. (\ref{M-action}) and (\ref%
{M-hat}) yield
\begin{equation}
(Z\,,{\overline{D}}_{\bar{\imath}}\bar{Z})=(Z_{0},Z_{i})\hat{M}\ ,
\end{equation}%
where $Z$ and $D_{i}Z$ in the l.h.s. are given by (\ref{Z2}) and (\ref{DaZ2}%
). Using (\ref{M-hat}), one finds
\begin{eqnarray}
Z &=&\frac{1}{2}(Z_{0}-i\lambda ^{i}Z_{i}\mathcal{V}^{-1/3})\ ;  \label{bZ}
\\
\overline{D}_{\bar{\imath}}\overline{Z} &=&\frac{1}{2}(\partial _{\bar{\imath%
}}KZ_{0}+\mathcal{V}^{-1/3}Z_{\bar{\imath}}+i\mathcal{V}^{-1/3}\lambda
^{j}Z_{j}\partial _{\bar{\imath}}K)\ .  \label{bDZ}
\end{eqnarray}%
In order to find the skew eigenvalues $z_{1},z_{2},z_{3},z_{4}$ in (\ref%
{skew-diag}), one needs the inverse metric, which in this case is factorized
as
\begin{equation}
g^{s\bar{s}}=-(s-\bar{s})^{2}\ ,\quad g^{t\bar{t}}=-(t-\bar{t})^{2}\ ,\quad
g^{u\bar{u}}=-(u-\bar{u})^{2}\ ,  \label{1}
\end{equation}%
as well as the purely imaginary \textit{Vielbein} (see App. \ref{App-1})
\begin{equation}
(g^{-1/2})_{\phantom s1}^{\bar{s}}=(s-\bar{s})\ ,\quad (g^{-1/2})_{\phantom %
t2}^{\bar{t}}=(t-\bar{t})\ ,\quad (g^{-1/2})_{\phantom u3}^{\bar{u}}=(u-\bar{%
u})\ ,  \label{2}
\end{equation}%
and the K\"{a}hler connection
\begin{equation}
\partial _{\bar{\imath}}K=\left( \frac{1}{s-\bar{s}},\frac{1}{t-\bar{t}},%
\frac{1}{u-\bar{u}}\right) ^{T}\ .  \label{3}
\end{equation}%
%
%
%
%
%
%
%
%
%
%
%
%
%
%
%
%
%
%
%
%
%
Using (\ref{1})-(\ref{3}) in (\ref{bZ})-(\ref{bDZ}), one obtains
\begin{eqnarray}
Z &=&\frac{1}{2}(Z_{0}-i\hat{\lambda}^{i}Z_{i})\ ; \\
\overline{D}_{\bar{s}}\bar{Z} &=&\frac{1}{2}\left( \frac{1}{s-\bar{s}}Z_{0}+%
\mathcal{V}^{-1/3}Z_{1}+i\mathcal{V}^{-1/3}\lambda ^{i}Z_{i}\frac{1}{s-\bar{s%
}}\right) \ ; \\
\overline{D}_{\bar{t}}\bar{Z} &=&\frac{1}{2}\left( \frac{1}{t-\bar{t}}Z_{0}+%
\mathcal{V}^{-1/3}Z_{2}+i\mathcal{V}^{-1/3}\lambda ^{i}Z_{i}\frac{1}{t-\bar{t%
}}\right) \ ; \\
\overline{D}_{\bar{u}}\bar{Z} &=&\frac{1}{2}\left( \frac{1}{u-\bar{u}}Z_{0}+%
\mathcal{V}^{-1/3}Z_{3}+i\mathcal{V}^{-1/3}\lambda ^{i}Z_{i}\frac{1}{u-\bar{u%
}}\right) \ .
\end{eqnarray}%
By recalling the definition $\lambda ^{i}\mathcal{V}^{-1/3}=\lambda
^{i}e^{-2\phi }\equiv \hat{\lambda}^{i}$ \texttt{(}\textit{cfr.} Sec. \ref%
{N=2}\texttt{)}, and defining
\begin{equation}
e_{1}\equiv \hat{\lambda}^{1}Z_{1}\ ,\quad e_{2}\equiv \hat{\lambda}%
^{2}Z_{2}\ ,\quad e_{3}\equiv \hat{\lambda}^{3}Z_{3}\ ,
\end{equation}%
one computes
\begin{eqnarray}
g^{s\bar{s}}\overline{D}_{\bar{s}}\bar{Z}D_{s}Z &=&\frac{1}{4}\left\vert
Z_{0}-i\hat{\lambda}^{1}Z_{1}+i\hat{\lambda}^{2}Z_{2}+i\hat{\lambda}%
^{3}Z_{3}\right\vert ^{2}=\frac{1}{4}\left\vert Z_{0}+i\left(
-e_{1}+e_{2}+e_{3}\right) \right\vert ^{2}; \\
g^{t\bar{t}}\overline{D}_{\bar{t}}\bar{Z}D_{t}Z &=&\frac{1}{4}\left\vert
Z_{0}+i\hat{\lambda}^{1}Z_{1}-i\hat{\lambda}^{2}Z_{2}+i\hat{\lambda}%
^{3}Z_{3}\right\vert ^{2}\ =\frac{1}{4}\left\vert Z_{0}+i\left(
e_{1}-e_{2}+e_{3}\right) \right\vert ^{2}; \\
g^{u\overline{u}}\overline{D}_{\overline{u}}\bar{Z}D_{u}Z &=&\frac{1}{4}%
\left\vert Z_{0}+i\hat{\lambda}^{1}Z_{1}+i\hat{\lambda}^{2}Z_{2}-i\hat{%
\lambda}^{3}Z_{3}\right\vert ^{2}\ =\frac{1}{4}\left\vert Z_{0}+i\left(
e_{1}+e_{2}-e_{3}\right) \right\vert ^{2},
\end{eqnarray}%
from which the entries of the $Z_{AB}$ matrix can be read off (in the
conventions of \textit{e.g.} (5.32) of \cite{Ceresole:2009vp})
\begin{eqnarray}
z_{1} &=&Z=\frac{i}{2}\left[ -(e_{1}+e_{2}+e_{3})-iZ_{0}\right] \ ,
\label{entriesZstu} \\
z_{2} &=&\overline{D}_{\bar{s}}\bar{Z}(g^{-1/2})_{\phantom s1}^{\bar{s}}=%
\frac{i}{2}\left( -iZ_{0}-\hat{\lambda}^{1}Z_{1}+\hat{\lambda}%
^{2}Z_{2}+\lambda ^{3}Z_{3}\right) =\nn \\
&=&\frac{i}{2}\left[ (e_{2}+e_{3}-e_{1})-iZ_{0}\right] \ , \\
z_{3} &=&\overline{D}_{\bar{t}}\bar{Z}(g^{-1/2})_{\phantom t2}^{\bar{t}}=%
\frac{i}{2}\left( -iZ_{0}+\hat{\lambda}^{1}Z_{1}-\hat{\lambda}%
^{2}Z_{2}+\lambda ^{3}Z_{3}\right) =\nn \\
&=&\frac{i}{2}\left[ (e_{1}+e_{3}-e_{2})-iZ_{0}\right] \ , \\
z_{4} &=&\overline{D}_{\bar{u}}\bar{Z}(g^{-1/2})_{\phantom u3}^{\bar{u}}=%
\frac{i}{2}\left( -iZ_{0}+\hat{\lambda}^{1}Z_{1}+\hat{\lambda}%
^{2}Z_{2}-\lambda ^{3}Z_{3}\right) =\nn \\
&=&\frac{i}{2}\left[ (e_{1}+e_{2}-e_{3})-iZ_{0}\right] \ .
\end{eqnarray}%
The $4D/5D$ covariant splitting is thus manifest in the following form of
the central charge matrix\footnote{$id_{n}$ denotes the $n\times n$ identity
matrix throughout.} \cite{Ceresole:2009id}
\begin{eqnarray}
Z_{AB} &=&\frac{i}{2}\epsilon \otimes \left[ -iZ_{0}~id_{4}+\left(
\begin{array}{cccc}
-e_{1}-e_{2}-e_{3} & 0 & 0 & 0 \\
0 & -e_{1}+e_{2}+e_{3} & 0 & 0 \\
0 & 0 & e_{1}-e_{2}+e_{3} & 0 \\
0 & 0 & 0 & e_{1}+e_{2}-e_{3}%
\end{array}%
\right) \right] \ .\nn  \label{ZABnostra} \\
&&
\end{eqnarray}%
This result, compared with formul\ae\ (3.2) of \cite{Ceresole:2009id},
explains the definition
\begin{equation*}
Z_{AB}=\frac{1}{2}\left( e_{AB}-iZ^{0}\Omega \right) \ ,
\end{equation*}%
in which $\Omega =\epsilon \otimes id_{4}$, given in Eq. (4.7) of the same
reference; notice that the overall phase $i$ is uninfluential.

\subsection{\label{Residual}Residual $U(1)^{3}$ Symmetry of the
Skew-Diagonal $Z_{AB}$}

The form of the central charge, as derived in the previous section, reflects
the more general structure of the $\mathbf{28}\rightarrow \mathbf{1}_{%
\mathbb{C}}+\mathbf{27}_{\mathbb{C}}$ decomposition of $SU(8)\supset USp(8)$
representation.

The central charge matrix for the $p^{0},q_{0}$ configuration in $N=8$
Supergravity has been given in \cite{Bossard:2009we}, in the same symplectic
frame. The reason why this is a suitable frame to study the non-BPS orbit is
related to the choice of orbit representative. The moduli space of the
non-BPS $p^{0},q_{0}$ solution is indeed the moduli space of the 5
dimensional theory, namely $E_{6(6)}/USp(8)$ . By solving a nonstandard
diagonalization problem, the authors of \cite{Bossard:2009we} identify the
fake-superpotential in the singlet of the axion-base decomposition of the
central charge matrix. However, the form of $Z_{AB}$ is unique up to $SU(8)$
transformations, and the choice of symplectic frame is not covariant with
respect to the action of $SU(8)$, since the singlet is not left invariant by
R-symmetry rotations.

Starting from the form of the central charge in \eqref{ZABnostra}, we look
for the transformation that rotates $Z_{AB}$ in such a way that the
transformed matrix can be identified with the one of \cite{Bossard:2009we}.
The goal is to determine the $SU(8)$ rotation in terms of the scalar fields,
and then read from the transformed singlet the explicit form of the fake
superpotential.

Because of the residual $USp(8)$ symmetry of the skew-diagonal central
charge \eqref{skew-diag}, we can restrict the analysis to the
transformations of $U(1)^3\subset SU(8)/USp(8)$.

\subsubsection{\label{D0-D6}The $\left( p^{0},q_{0}\right) $ Configuration}

In the non-BPS $\left( p^{0},q_{0}\right) $ charge configuration
(corresponding to $D0-D6$ in Type II language), the dressed charges of the $%
N=8$ theory read (\ref{cariche}) 
\begin{eqnarray}
Z_{0} &=&\frac{1}{\sqrt{2}}\left( e^{-3\phi }q_{0}+e^{-3\phi
}p^{0}a_{1}a_{2}a_{3}+ie^{3\phi }p^{0}\right) \ ; \\
Z_{i} &=&\frac{1}{\sqrt{2}}p^{0}\left[ e^{-\phi }\left(
\begin{array}{c}
\hat{\lambda}^{1}a^{2}a^{3} \\
\hat{\lambda}^{2}a^{1}a^{3} \\
\hat{\lambda}^{3}a^{1}a^{2}%
\end{array}%
\right) -ie^{\phi }\left(
\begin{array}{c}
\frac{a^{1}}{\hat{\lambda}^{1}} \\
\frac{a^{2}}{\hat{\lambda}^{2}} \\
\frac{a^{3}}{\hat{\lambda}^{3}}%
\end{array}%
\right) \right] \ .
\end{eqnarray}%
Thus, the $N=8$ skew-diagonal $Z_{AB}$ (\ref{skew-diag}) in the $\left(
p^{0},q_{0}\right) $ charge configuration can then be written as
\begin{eqnarray}
Z_{AB}^{(p_{0},q_{0})} &=&\frac{1}{2\sqrt{2}}\epsilon \otimes \left[
(e^{-3\phi }q_{0}+\alpha _{1}\alpha _{2}\alpha _{3}\,p^{0}e^{3\phi
}+ip^{0}e^{3\phi })\left(
\begin{array}{cccc}
1 & 0 & 0 & 0 \\
0 & 1 & 0 & 0 \\
0 & 0 & 1 & 0 \\
0 & 0 & 0 & 1%
\end{array}%
\right) +\right.   \label{Zaxp0q0} \\
&+&\left. p^{0}e^{3\phi }\left(
\begin{array}{cccc}
-(\alpha _{1}+\alpha _{2}+\alpha _{3}) & 0 & 0 & 0 \\
0 & -\alpha _{1}+\alpha _{2}+\alpha _{3} & 0 & 0 \\
0 & 0 & \alpha _{1}-\alpha _{2}+\alpha _{3} & 0 \\
0 & 0 & 0 & \alpha _{1}+\alpha _{2}-\alpha _{3}%
\end{array}%
\right) \right. +\nn \\
&+&\left. p^{0}ie^{3\phi }(\alpha _{1}\alpha _{2}\alpha _{3})\left(
\begin{array}{cccc}
-(\frac{1}{\alpha _{1}}+\frac{1}{\alpha _{2}}+\frac{1}{\alpha _{3}}) & 0 & 0
& 0 \\
0 & (-\frac{1}{\alpha _{1}}+\frac{1}{\alpha _{2}}+\frac{1}{\alpha _{3}}) & 0
& 0 \\
0 & 0 & (\frac{1}{\alpha _{1}}-\frac{1}{\alpha _{2}}+\frac{1}{\alpha _{3}})
& 0 \\
0 & 0 & 0 & (\frac{1}{\alpha _{1}}+\frac{1}{\alpha _{2}}-\frac{1}{\alpha _{3}%
})%
\end{array}%
\right) \right] \ ,\nn \\
&&
\end{eqnarray}%
where $\alpha ^{i}\equiv a^{i}/\lambda ^{i}$ is the axion/dilaton ratio,
with $\lambda ^{i}=e^{2\phi }\hat{\lambda}^{i}$, and $\hat{\lambda}^{1}\hat{%
\lambda}^{2}\hat{\lambda}^{3}=1$. When $a^{i}=0$, one recovers the KK
solution studied in \cite{Ceresole:2009id}.

To proceed further, it is convenient to define the following quantities:
\begin{eqnarray}
Y_{0} &=&\frac{1}{\sqrt{2}}(q_{0}\,e^{-3\phi }+\alpha _{1}\alpha _{2}\alpha
_{3}\,p^{0}e^{3\phi})+\frac{i}{\sqrt{2}}p^{0}e^{3\phi }; \\
Y_{i} &=&-\frac{1}{\sqrt{2}}p^{0}e^{3\phi}\left( \alpha _{i}+\frac{i}{2}%
\left\vert \epsilon _{ijk}\right\vert \alpha _{j}\alpha _{k}\right) ,
\end{eqnarray}%
and%
\begin{equation}
\sigma _{3}=\left(
\begin{array}{cc}
1 & 0 \\
0 & -1%
\end{array}%
\right) \ .
\end{equation}
We can write
\begin{eqnarray}
id_{2}\otimes id_{2} &=&id_{4}\ ,\qquad id_{2}\otimes \sigma _{3}=\left(
\begin{array}{cccc}
\ 1 & \ 0 &  &  \\
\ 0 & \ 1 &  &  \\
&  & -1 & \ 0 \\
&  & \ 0 & -1%
\end{array}%
\right) \ ,\qquad \sigma _{3}\otimes id_{2}=\left(
\begin{array}{cccc}
\ 1 & \ 0 &  &  \\
\ 0 & \ -1 &  &  \\
&  & 1 & \ 0 \\
&  & \ 0 & -1%
\end{array}%
\right) ; \nn  \label{coolMatr} \\
\sigma _{3}\otimes \sigma _{3} &=&\left(
\begin{array}{cccc}
\ 1 & \ 0 &  &  \\
\ 0 & \ -1 &  &  \\
&  & -1 & \ 0 \\
&  & \ 0 & 1%
\end{array}%
\right) .
\end{eqnarray}%
Thus, by recalling (\ref{Zaxp0q0}), $Z_{AB}$ can be decomposed as
\begin{equation}
Z_{AB}^{(p^{0},q_{0})}\equiv Z_{AB}(Y_{0},Y_{i})=\frac{1}{2}\epsilon \otimes %
\left[ Y_{0}\ id_{4}+Y_{1}\ id_{2}\otimes \sigma _{3}+Y_{2}\ \sigma
_{3}\otimes id_{2}+Y_{3}\ \sigma _{3}\otimes \sigma _{3}\right] .
\label{Zpar}
\end{equation}

This parametrization of the central charge matrix will allow us to perform
the necessary rotation to identify the fake superpotential.

\subsubsection{\label{U(1)^3}$U(1)^{3}$}

The matrix $Z_{AB}$ (\ref{Zpar}) has a residual $U(1)^{3}\subset
SU(8)/USp(8) $ symmetry. More precisely, $U(1)^{3}$ can be considered as the
Cartan subalgebra of the symmetric, rank-3 compact manifold $SU(8)/USp(8)$
(dim$_{\mathbb{R}}=27$); indeed, $U(1)^{3}$-transformations do not generate
off-diagonal elements, and they leave the skew-diagonal form of $Z_{AB}$
invariant. We choose to parametrize such a $U(1)^{3}$ matrix as a $4\times 4$
matrix acting on the diagonal part of $Z_{AB}$, namely ($\chi _{i}\in
\mathbb{R}$)
\begin{equation}
\mathcal{U}\equiv \left(
\begin{array}{cccc}
e^{-i(\chi _{1}+\chi _{2}+\chi _{3})} &  &  &  \\
& e^{i(-\chi _{1}+\chi _{2}+\chi _{3})} &  &  \\
&  & e^{i(\chi _{1}-\chi _{2}+\chi _{3})} &  \\
&  &  & e^{i(\chi _{1}+\chi _{2}-\chi _{3})}%
\end{array}%
\right) \in U(1)^{3}\subset SU(8)/USp(8)\ .
\end{equation}%
Note that, consistently, the sum of the four diagonal phases vanishes.
Therefore, by the exponential mapping, one obtains
\begin{equation}
\mathcal{U}=\exp \left(
\begin{array}{cccc}
{-i(\chi _{1}+\chi _{2}+\chi _{3})} &  &  &  \\
& {i(-\chi _{1}+\chi _{2}+\chi _{3})} &  &  \\
&  & {i(\chi _{1}-\chi _{2}+\chi _{3})} &  \\
&  &  & {i(\chi _{1}+\chi _{2}-\chi _{3})}%
\end{array}%
\right) ,  \label{U-exp}
\end{equation}%
which, analogously to $Z_{AB}$ (\ref{Zpar}), enjoys the following
decomposition :
\begin{eqnarray}
\mathcal{U} &=&\exp \left[ -i(\chi _{1}\ \ id_{2}\otimes \sigma _{3}+\chi
_{2}\ \sigma _{3}\otimes id_{2}+\chi _{3}\ \sigma _{3}\otimes \sigma _{3})%
\right] =\nn \\
&=&\exp \left[ -i\chi _{1}\ \ id_{2}\otimes \sigma _{3}\right] \cdot \exp %
\left[ -i\chi _{2}\ \sigma _{3}\otimes id_{2}\right] \cdot \exp \left[
-i\chi _{3}\ \sigma _{3}\otimes \sigma _{3}\right] =\nn \\
&=&\mathcal{U}_{1}\cdot \mathcal{U}_{2}\cdot \mathcal{U}_{3}  \label{U-exp-2}
\end{eqnarray}%
where all matrices are reciprocally commuting.

Under $U(1)^{3}$ (\ref{U-exp}), $Z_{AB}$ (\ref{Zpar}) transforms as
\begin{equation}
Z_{AB}\rightarrow \mathcal{U}Z_{AB}\mathcal{U}^{T}\equiv \mathcal{U}%
^{2}Z_{AB}.
\end{equation}%
Without loss of generality, one can therefore just redefine the $\chi _{i}$%
's by a factor of $2$, and consider the transformation%
\begin{equation}
Z_{AB}\rightarrow \mathcal{U}Z_{AB}.
\end{equation}%
Each single $\mathcal{U}_{i}$ actually reads%
\begin{equation}
\begin{array}{l}
\mathcal{U}_{1}=\exp \left[ -i\chi _{1}\ \ id_{2}\otimes \sigma _{3}\right]
=\cos \chi _{1}\ id_{4}-i\sin \chi _{1}\ id_{2}\otimes \sigma _{3}; \\
\mathcal{U}_{2}=\exp \left[ -i\chi _{2}\ \sigma _{3}\otimes id_{2}\right]
=\cos \chi _{2}\ id_{4}-i\sin \chi _{2}\ \sigma _{3}\otimes id_{2}; \\
\mathcal{U}_{3}=\exp \left[ -i\chi _{3}\ \sigma _{3}\otimes \sigma _{3}%
\right] =\cos \chi _{3}\ id_{4}-i\sin \chi _{3}\sigma _{3}\otimes \sigma
_{3},%
\end{array}%
\end{equation}
and induces the following transformation on $Z_{AB}$ (\ref{Zpar}):%
\begin{equation}
\begin{array}{l}
\mathcal{U}_{1}\,Z_{AB}\rightarrow \cos \chi _{1}\ Z_{AB}-i\sin \chi _{1}\
Z_{AB}\cdot id_{2}\otimes \sigma _{3}; \\
\mathcal{U}_{2}\,Z_{AB}\rightarrow \cos \chi _{2}\ Z_{AB}-i\sin \chi _{2}\
Z_{AB}\cdot \sigma _{3}\otimes id_{2}; \\
\mathcal{U}_{3}\,Z_{AB}\rightarrow \cos \chi _{3}\ Z_{AB}-i\sin \chi _{3}\
Z_{AB}\cdot \sigma _{3}\otimes \sigma _{3}.%
\end{array}%
\end{equation}
Consequently, $\mathcal{U}$ (\ref{U-exp-2}) has a well defined action on the
coefficients of the matrices (\ref{coolMatr}); for example, by acting with
only $\mathcal{U}_{1}$ gives rise to the following transformations of $Y_{0}$
and $Y_{i}$'s:%
\begin{equation}
\begin{array}{l}
Y_{0}\rightarrow \gamma _{0}\equiv \cos \chi _{1}\ Y_{0}-i\sin \chi _{1}\
Y_{1}; \\
Y_{1}\rightarrow \gamma _{1}\equiv \cos \chi _{1}\ Y_{1}-i\sin \chi _{1}\
Y_{0}; \\
Y_{2}\rightarrow \gamma _{2}\equiv \cos \chi _{1}\ Y_{2}-i\sin \chi _{1}\
Y_{3}; \\
Y_{3}\rightarrow \gamma _{3}\equiv \cos \chi _{1}\ Y_{3}-i\sin \chi _{1}\
Y_{2},%
\end{array}
\label{gamma}
\end{equation}
such that the $\mathcal{U}_{1}$-transformed central charge matrix (\ref{Zpar}%
) can be rewritten as
\begin{equation}
Z_{AB}(Y_{0},Y_{i})\rightarrow \mathcal{U}_{1}Z_{AB}(Y_{0},Y_{i})=Z_{AB}(%
\gamma _{0},\gamma _{i})\ .
\end{equation}%
The complete action of $\mathcal{U}$ (\ref{U-exp-2}) on (\ref{Zpar}) reads
\begin{equation}
Z_{AB}(Y_{0},Y_{i})\rightarrow Z_{AB}(\zeta _{0},\zeta _{i})=\mathcal{U}%
_{3}\,\mathcal{U}_{2}\,\mathcal{U}_{1}Z_{AB}(Y_{0},Y_{i})\ ,  \label{trtr}
\end{equation}%
where the $\zeta _{I}$'s are defined as%
\begin{equation}
\begin{array}{l}
\zeta _{0}\equiv A\ Y_{0}+B\ Y_{1}+C\ Y_{2}+D\ Y_{3}; \\
\zeta _{1}\equiv B\ Y_{0}+AY_{1}+D\ Y_{2}+C\ Y_{3}; \\
\zeta _{2}\equiv C\ Y_{0}+D\ Y_{1}+A\ Y_{2}+B\ Y_{3}; \\
\zeta _{3}\equiv D\ Y_{0}+C\ Y_{1}+B\ Y_{2}+A\ Y_{3},%
\end{array}
\label{system1}
\end{equation}
with ($c_{i}\equiv \cos \chi _{i}$, $s_{i}\equiv \sin \chi _{i}$)%
\begin{equation}
\begin{array}{l}
A\equiv (c_{1}c_{2}c_{3}-is_{1}s_{2}s_{3}); \\
B\equiv (-c_{1}s_{2}s_{3}+is_{1}c_{2}c_{3}); \\
C\equiv (-s_{1}c_{2}s_{3}+ic_{1}s_{2}c_{3}); \\
D\equiv (-s_{1}s_{2}c_{3}+ic_{1}c_{2}s_{3}).%
\end{array}
\label{system2}
\end{equation}

Within the same $\left( p^{0},q_{0}\right) $ axionful charge configuration,
it is interesting to compare the $U(1)^{3}$-transformed $Z_{AB}$ (\ref{trtr}%
)-(\ref{system2}) with the \textquotedblleft non-standard" skew-diagonalized
$Z_{AB}^{(BMP)}$ obtained by Bossard, Michel and Pioline (BMP) in
\cite{Bossard:2009we}
\begin{eqnarray}
Z_{AB}^{(BMP)} &=&\frac{1}{2}\epsilon \otimes \left[ i(e^{i(\alpha -\pi
/4)}+\sin 2\alpha e^{-i(\alpha -\pi /4)})\rho \left(
\begin{array}{cccc}
1 & 0 & 0 & 0 \\
0 & 1 & 0 & 0 \\
0 & 0 & 1 & 0 \\
0 & 0 & 0 & 1%
\end{array}%
\right) \right. \nn \\
&&\left. \hspace{1.5cm}+e^{-i(\alpha -\pi /4)}\left(
\begin{array}{cccc}
\xi _{1}+\xi _{2}+\xi _{3} & 0 & 0 & 0 \\
0 & -\xi _{1} & 0 & 0 \\
0 & 0 & -\xi _{2} & 0 \\
0 & 0 & 0 & -\xi _{3}%
\end{array}%
\right) \right] \ ,\nn  \label{Zalpha} \\
&&
\end{eqnarray}%
which can equivalently be recast in the following form:
\begin{eqnarray}
Z_{AB}^{(BMP)} &=&\frac{1}{2}\epsilon \otimes \left[ i(e^{i\eta }+\cos 2\eta
\,e^{-i\eta })\rho \,id_{4}+\right. \nn  \label{ZBMP} \\
&&\ \ \left. +e^{-i\eta }\left(
\begin{array}{cccc}
-\mu _{1}-\mu _{2}-\mu _{3} & 0 & 0 & 0 \\
0 & -\mu _{1}+\mu _{2}+\mu _{3} & 0 & 0 \\
0 & 0 & \mu _{1}-\mu _{2}+\mu _{3} & 0 \\
0 & 0 & 0 & \mu _{1}+\mu _{2}-\mu _{3}%
\end{array}%
\right) \right] =\nn \\
&=&\frac{1}{2}\epsilon \otimes \left[ \mu _{0}\ id_{4}-e^{-i\eta }\mu _{1}\
id_{2}\otimes \sigma _{3}-e^{-i\eta }\mu _{2}\ \sigma _{3}\otimes
id_{2}-e^{-i\eta }\mu _{3}\ \sigma _{3}\otimes \sigma _{3}\right] \nn \\
&&
\end{eqnarray}%
by introducing the quantities:%
\begin{equation}
\begin{array}{l}
\mu _{0}\equiv i(e^{i\eta }+\cos 2\eta \,e^{-i\eta })\rho \ ,\quad \eta
\equiv \alpha -\frac{\pi }{4}\ , \\
\xi _{1}\equiv \mu _{1}-\mu _{2}-\mu _{3}\ ,\quad \xi _{2}\equiv -\mu
_{1}+\mu _{2}-\mu _{3}\ ,\quad \xi _{3}\equiv -\mu _{1}-\mu _{2}+\mu _{3}.%
\end{array}
\label{defs-1}
\end{equation}%
By comparing (\ref{Zpar}) and (\ref{ZBMP}), in order to match (\ref{Zalpha})
with (\ref{trtr})-(\ref{system2}), a transformation $\mathcal{U\in }U(1)^{3}$
should be found, such that
\begin{equation}
Y_{0}\rightarrow \zeta _{0}=\mu _{0}\ ,\quad Y_{i}\rightarrow \zeta
_{i}=-e^{-i\eta }\mu _{i}\ ,\qquad i=1,2,3\ .  \label{system3}
\end{equation}%
This amounts to solving the system composed by (\ref{system1})-(\ref{system2}%
) and (\ref{defs-1})-(\ref{system3}). For simplicity's sake, we will here
confine ourselves to solve such a system within the \textquotedblleft $t^{3}$%
-degeneration" of the formalism under consideration, which amounts to
choosing three equal phases $\chi _{i}$'s, corresponding to the diagonal $%
U(1)_{diag}$ inside $U(1)^{3}$.

\subsubsection{\label{T^3-Model}$t^{3}$ model}

As mentioned, at the level of $\mathcal{U}$-transformation, the
\textquotedblleft degeneration" procedure from $stu$ to $t^{3}$ model
amounts to identifying%
\begin{equation}
\chi _{1}=\chi _{2}=\chi _{3}\equiv \chi .
\end{equation}%
This corresponds to considering the action of $U(1)_{diag}\subset
U(1)^{3}\subset SU(8)/USp(8)$, such that (recall (\ref{U-exp-2}))
\begin{equation}
\mathcal{U}=\mathcal{U}_{1}\,\cdot \mathcal{U}_{2}\,\cdot \mathcal{U}%
_{3}\equiv \mathbb{U}_{diag}=\exp \left(
\begin{array}{cccc}
{-3i\chi } &  &  &  \\
& {i\chi } &  &  \\
&  & {i\chi } &  \\
&  &  & {i\chi }%
\end{array}%
\right) .  \label{U-diag}
\end{equation}

The central charge matrix given by (\ref{Zaxp0q0}) and (\ref{Zpar}) thus
acquires the following structure\footnote{%
In order to simplify the computation, we will henceforth choose $p^{0}>0$
and $q_{0}>0$. This does not imply any loss of generality, since all other
sign choices are related to this by a duality rotation along the non-BPS ($%
Z_{H}\neq 0$) charge orbit of the $stu$ model
.}:
\begin{eqnarray}
Z_{AB}^{(p^{0},q_{0}),t^{3}} &=&\frac{1}{2\sqrt{2}}\epsilon \otimes \left[
\left( e^{-3\phi }q_{0}+p^{0}e^{3\phi }(i+\alpha ^{3})\right)
\,id_{4}+p^{0}\alpha e^{3\phi }(1+i\alpha )\left(
\begin{array}{cccc}
-3 & 0 & 0 & 0 \\
0 & 1 & 0 & 0 \\
0 & 0 & 1 & 0 \\
0 & 0 & 0 & 1%
\end{array}%
\right) \right] \nn  \label{Zt3} \\
&=&\frac{1}{2}\epsilon \otimes \left[ Y_{0}\ id_{4}+Y\ \left( id_{2}\otimes
\sigma _{3}+\sigma _{3}\otimes id_{2}+\sigma _{3}\otimes \sigma _{3}\right) %
\right] \ ,
\end{eqnarray}%
where here ($\alpha _{1}=\alpha _{2}=\alpha _{3}\equiv \alpha $)%
\begin{equation}
\begin{array}{l}
Y_{0}\equiv \frac{1}{\sqrt{2}}\left( e^{-3\phi }q_{0}+p^{0}e^{3\phi
}(i+\alpha ^{3})\right) \ ; \\
Y\equiv -\frac{1}{\sqrt{2}}\left( p^{0}e^{3\phi }\alpha (1+i\alpha )\right)
\ .%
\end{array}
\label{Ys}
\end{equation}%
%
%
%
%
%
%
%

On the other hand, the consistent \textquotedblleft $t^{3}$-degeneration" of
the central charge matrix (\ref{ZBMP})-(\ref{defs-1}) reads
\begin{eqnarray}
Z_{AB}^{(BMP),t^{3}} &=&\frac{1}{2}\epsilon \otimes \left[ i(e^{i\eta }+\cos
2\eta \,e^{-i\eta })\rho \ id_{4}+e^{-i\eta }\mu \left(
\begin{array}{cccc}
-3 & 0 & 0 & 0 \\
0 & 1 & 0 & 0 \\
0 & 0 & 1 & 0 \\
0 & 0 & 0 & 1%
\end{array}%
\right) \right] =\nn  \label{ZBMPt3} \\
&=&\frac{1}{2}\epsilon \otimes \left[ \mu _{0}\ id_{4}-e^{-i\eta }\mu (\
id_{2}\otimes \sigma _{3}+\sigma _{3}\otimes id_{2}+\sigma _{3}\otimes
\sigma _{3})\right] ,
\end{eqnarray}%
where
\begin{equation}
\mu _{0}\equiv i(e^{i\eta }+\cos 2\eta \,e^{-i\eta })\rho \ ,\quad \eta
\equiv \alpha -\frac{\pi }{4}\quad \mu _{1}=\mu _{2}=\mu _{3}\equiv \mu
\equiv -\xi _{1}=-\xi _{2}=-\xi _{3}\ .  \label{def-def}
\end{equation}%
We notice that, by denoting $\eta _{0}$ the phase of $\mu _{0}$, it holds
that
\begin{equation}
\tan \eta _{0}=-\frac{1}{\left( \tan \eta \right) ^{3}}\ .
\label{conditionChi}
\end{equation}

Thus, in order to match (\ref{Zt3})-(\ref{Ys}) with (\ref{ZBMPt3})-(\ref%
{def-def}), a phase $\chi $ should be determined such that it rotates the
relevant quantities as follows ($\zeta _{1}=\zeta _{2}=\zeta _{3}\equiv
\zeta $, $Y_{1}=Y_{2}=Y_{3}\equiv Y$)
\begin{equation}
Y_{0}\rightarrow \zeta _{0}=\mu _{0}\ ,\quad Y\rightarrow \zeta =-e^{-i\eta
}\mu \ .  \label{t-t}
\end{equation}%
From the \textquotedblleft $t^{3}$-degeneration" of (\ref{system1}), one
gets
\begin{eqnarray}
\zeta _{0} &=&AY_{0}+3BY;  \label{zeta-1} \\
\zeta  &=&(A+2B)Y+BY_{0}.  \label{zeta-2}
\end{eqnarray}%
However, now $A$ and $B$ respectively simplifies down to%
\begin{equation}
A\equiv B+e^{-i\chi },~~B\equiv \frac{i}{2}\sin (2\chi )e^{i\chi },
\end{equation}%
thus allowing for the following re-writing of (\ref{zeta-1})-(\ref{zeta-2}):
\begin{equation}
\begin{array}{l}
\zeta _{0}=e^{-i\chi }Y_{0}+\frac{i}{2}\sin (2\chi )e^{i\chi }(3Y+Y_{0}); \\
\zeta =e^{-i\chi }Y+\frac{i}{2}\sin (2\chi )e^{i\chi }(3Y+Y_{0})\ .%
\end{array}
\label{zetas}
\end{equation}%
The action of $U(1)_{diag}\subset U(1)^{3}$ implies that
\begin{equation}
e^{i\chi }=\frac{Y-Y_{0}}{\zeta -\zeta _{0}}.
\end{equation}%
As pointed out above, in order to match (\ref{Zt3})-(\ref{Ys}) with (\ref%
{ZBMPt3})-(\ref{def-def}), we are interested in finding the phases of these
parameters in terms of $\chi $ entering (\ref{U-diag}). Therefore, we can
solve for $\tan \chi $, as we read from (\ref{conditionChi}) and (\ref{t-t}%
):
\begin{equation}
\eta _{0}=\tan \psi (\zeta _{0})=\frac{1}{\left[ \tan \psi (\zeta )\right]
^{3}},  \label{tosolve}
\end{equation}%
where $\psi (\zeta _{0})$ and $\psi (\zeta )$ respectively denote the phases
of $\zeta _{0}$ and $\zeta $.

From (\ref{zetas}), one obtains
\begin{eqnarray}
\tan \psi (\zeta _{0}) &=&\frac{1}{\tau ^{3}}\frac{Y_{0I}-\tau
^{3}Y_{0R}-3\tau ^{2}Y_{I}+3\tau Y_{R}}{Y_{0I}+\frac{1}{\tau ^{3}}Y_{0R}-%
\frac{3}{\tau ^{2}}Y_{I}-\frac{3}{\tau }Y_{R}}\ ;  \label{res-1} \\
\tan \psi (\zeta ) &=&\frac{1}{\tau ^{3}}\frac{Y_{I}-\tau ^{3}Y_{R}-\tau
^{2}(2Y_{I}+Y_{0I})+\tau (2Y_{R}+Y_{0R})}{Y_{I}+\frac{1}{\tau ^{3}}Y_{0R}-%
\frac{1}{\tau ^{2}}(2Y_{I}+Y_{0I})-\frac{1}{\tau }(2Y_{R}+Y_{0R})}\ ,
\label{res-2}
\end{eqnarray}%
where%
\begin{equation}
Y\equiv Y_{R}+iY_{I},~~Y_{0}\equiv Y_{0R}+iY_{0I},~~\tau \equiv \tan \chi .
\end{equation}
In order to find $\tau $ in terms of $\alpha ,p^{0},q_{0}$, one needs to
solve (\ref{tosolve}), which in virtue of (\ref{res-1})-(\ref{res-2}) can be
made explicit as
\begin{equation}
\frac{Y_{0I}-\tau ^{3}Y_{0R}-3\tau ^{2}Y_{I}+3\tau Y_{R}}{Y_{0I}+\frac{1}{%
\tau ^{3}}Y_{0R}-\frac{3}{\tau ^{2}}Y_{I}-\frac{3}{\tau }Y_{R}}=\tau ^{12}%
\left[ \frac{Y_{I}+\frac{1}{\tau ^{3}}Y_{0R}-\frac{1}{\tau ^{2}}%
(2Y_{I}+Y_{0I})-\frac{1}{\tau }(2Y_{R}+Y_{0R})}{Y_{I}-\tau ^{3}Y_{R}-\tau
^{2}(2Y_{I}+Y_{0I})+\tau (2Y_{R}+Y_{0R})}\right] ^{3}.  \label{tosolve-2}
\end{equation}

Further simplifications are possible. Indeed, by recalling (\ref{Ys}), the
dependence of (\ref{res-1})-(\ref{res-2}) on $\alpha ,e^{\phi },p^{0},q^{0}$
can be made manifest:
\begin{eqnarray}
\eta _{0} &=&\tan \psi (\zeta _{0})=\frac{-q_{0}\tan \chi ^{3}+p^{0}e^{6\phi
}(1-\tan \chi \ \alpha )^{3}}{q_{0}+p^{0}e^{6\phi }(\tan \chi +\alpha )^{3}}
\notag \\
&=&-\tan \chi ^{3}\frac{1-\frac{p^{0}}{q_{0}}e^{6\phi }(\frac{1}{\tan \chi }%
-\alpha )^{3}}{1+\frac{p^{0}}{q_{0}}e^{6\phi }(\tan \chi +\alpha )^{3}};
\label{tanzs-1} \\
\tan \psi (\zeta ) &=&\frac{-q_{0}\tan \chi +p^{0}e^{6\phi }(\tan \chi
+\alpha )^{2}(1-\tan \chi \ \alpha )}{q_{0}\tan \chi ^{2}+p^{0}e^{6\phi
}(\tan \chi +\alpha )(1-\tan \chi \ \alpha )^{2}}\   \notag \\
&=&-\frac{1}{\tan \chi }\ \frac{1-\frac{p^{0}}{q_{0}}e^{6\phi }(\tan \chi
+\alpha )^{2}(\frac{1}{\tan \chi }-\alpha )}{1+\frac{p^{0}}{q_{0}}e^{6\phi
}(\tan \chi +\alpha )(\frac{1}{\tan \chi }-\alpha )^{2}}.  \label{tanzs-2}
\end{eqnarray}%
As a consequence, (\ref{tosolve}) can be recast as
\begin{eqnarray}
\frac{1-x^{3}}{1+y^{3}} &=&\frac{(1+x^{2}y)^{3}}{(1-x\,y^{2})^{3}}\ ,
\label{relazfasi1} \\
x &\equiv &\left( \frac{p^{0}}{q_{0}}\right) ^{1/3}e^{2\phi }\left( \frac{1}{%
\tan \chi }-\alpha \right) \ ,\qquad y\equiv \left( \frac{p^{0}}{q_{0}}%
\right) ^{1/3}e^{2\phi }\left( \tan \chi +\alpha \right) ,
\end{eqnarray}%
and therefore solved for
\begin{equation}
x=y\qquad \text{or}\quad x\neq y\ \ ,\ \ xy=-1\ .  \label{solut1}
\end{equation}%
For real values of $\tan \chi $ the case $x=y$ is not allowed, so one is
left with
\begin{equation}
xy=-1\quad \Rightarrow \quad \left( \frac{p^{0}}{q_{0}}\right)
^{2/3}e^{4\phi }\left[ 1-\alpha ^{2}+\frac{2\alpha }{\tan 2\chi }\right]
=-1\ .
\end{equation}

Thus, the angle $\chi $, which provides the $U(1)_{diag}$-rotation between
the skew-eigenvalues of (\ref{Zt3}) and (\ref{ZBMPt3}), is given by
\begin{eqnarray}
\tan \chi &=&\frac{1}{2\nu ^{2/3}\alpha }\left( (1-\nu ^{2/3}(\alpha
^{2}+1))\pm \sqrt{(1-\nu ^{2/3}(\alpha ^{2}+1))^{2}+4\nu ^{2/3}}\right) \ ,
\label{chisol} \\
\nu &\equiv &(p^{0}/q_{0})e^{6\phi }.
\end{eqnarray}
For later convenience we explicite here the expression for $\chi$
\begin{eqnarray}  \label{chi1}
\chi&=&-\frac12\arctan\left[\frac{2\alpha} {(\frac{q_0}{p^0}%
)^{2/3}e^{4\phi}-1+\alpha^2}\right]\ ;
\end{eqnarray}
we also recall the choice of $q_0>0$, $p^0>0$, in our computation.

\subsubsection{Duality Invariants}

One can also relate the parameters entering the solution (\ref{chisol}) to
the duality invariants $\mathcal{I}_{4}$, $i_{1},i_{2}$ and $i_{3}$ defined
\textit{e.g.} in \cite{CFMZ1}. Using the relations (3.6)-(3.10) of \cite%
{Ceresole:2009iy}, one finds
\begin{eqnarray}
\alpha &=&\frac{b}{3\sqrt{-\mathcal{I}_{4}}}\ ; \\
(q^{0})^{2}e^{-6\phi } &=&\frac{1}{(-\mathcal{I}_{4})}\Big(4i_{3}\sqrt{-%
\mathcal{I}_{4}}\pm \sqrt{b^{6}-\mathcal{I}_{4}(3b^{4}+16i_{3}^{3})+3b^{2}(-%
\mathcal{I}_{4})^{2}-\mathcal{I}_{4}^{3}}\Big),
\end{eqnarray}%
where $i_{2}=b+3i_{1}$, and the \textquotedblleft $\pm $" choice has to be
consistent with the positivity of $e^{6\phi }$. We notice that $\alpha $ is
a duality invariant quantity by itself, as well as the combinations $%
q_{0}e^{-3\phi }$ and $p^{0}e^{3\phi }$ (recall $\sqrt{-\mathcal{I}_{4}}%
=p^{0}q_{0}$). Thus, the expression (\ref{chisol}) is explicitly duality
invariant.

\subsection{\label{Recover}Recovering the non-BPS Fake Superpotential}

In \cite{Bossard:2009we} it is shown that the non-BPS fake superpotential is
given by
\begin{equation}
W=2\rho ,
\end{equation}%
where $\rho $ enters the expression (\ref{ZBMP}). From the same equation,
one can also write $\mu _{0}$ as
\begin{equation}
\mu _{0}=2\rho (-\sin \eta ^{3}+i\cos \eta ^{3})\ ,
\end{equation}%
thus yielding
\begin{equation}
W=2\rho =\frac{\mathrm{Im}\mu _{0}}{\cos \eta ^{3}}\equiv \frac{\mathrm{Im}%
\zeta _{0}}{\cos \eta ^{3}}\ .  \label{jazz}
\end{equation}%
Moreover, (\ref{Ys}) and (\ref{zetas}) imply
\begin{eqnarray}
\mathrm{Im}\zeta _{0} &=&-\frac{1}{\sqrt{2}}e^{-3\phi }\cos \chi ^{3}\left(
q_{0}\tan \chi ^{3}-e^{6\phi }p^{0}(1-\tan \chi \,\alpha )^{3}\right) =\nn \\
&=&-\frac{1}{\sqrt{2}}e^{-3\phi }q_{0}\sin \chi ^{3}\left( 1-\nu \left(
\frac{1}{\tan \chi }-\alpha \right) ^{3}\right) \ .
\end{eqnarray}%
By using%
\begin{equation}
\sin \chi ^{3}=\frac{\tan \chi ^{3}}{(1+\tan \chi ^{2})^{3/2}}\ ,\quad \frac{%
1}{\cos \eta ^{3}}=(1+\tan \phi (\zeta )^{2})^{3/2}\ ,
\end{equation}%
and (\ref{tanzs-1})-(\ref{tanzs-2}), (\ref{tosolve}) and (\ref{solut1})
yield
\begin{equation}
\nu ^{2/3}(1/\tau -\alpha )(\tau +\alpha )=-1\ ,
\end{equation}%
and one can rewrite
\begin{eqnarray}
\tan \psi (\zeta ) &=&-\frac{1}{\tau }\frac{1+\nu ^{1/3}(\tau +\alpha )}{%
1-\nu ^{1/3}(1/\tau -\alpha )}\nn \\
&\Downarrow &\nn \\
\frac{1}{\cos \phi (\zeta )^{3}} &=&\frac{1}{\tau ^{3}}\frac{\left( (1-\nu
^{1/3}(1/\tau -\alpha ))^{2}+(1+\nu ^{1/3}(\tau +\alpha ))^{2}\right) ^{3/2}%
}{(1-\nu ^{1/3}(1/\tau -\alpha ))^{3}}=\nn \\
&=&\frac{(1+\tau ^{2})^{3/2}}{\tau ^{3}}\frac{(1+2\alpha \nu ^{1/3}+\nu
^{2/3}(\alpha ^{2}+1))^{3/2}}{(1-\nu ^{1/3}(1/\tau -\alpha ))^{3}}; \\
&&  \notag \\
\mathrm{Im}\zeta _{0} &=&-q_{0}e^{-3\phi }\frac{\tau ^{3}}{(1+\tau
^{2})^{3/2}}\left( 1-\nu \left( 1/{\tau }-\alpha \right) ^{3}\right) \ .
\end{eqnarray}%
Therefore, the non-BPS fake superpotential $W$ (\ref{jazz}) is given by
\begin{equation}
W=-\frac{1}{\sqrt{2}}q_{0}e^{-3\phi }\frac{\left( 1-\nu \left( 1/{\tau }%
-\alpha \right) ^{3}\right) }{\left( 1-\nu ^{1/3}\left( 1/{\tau }-\alpha
\right) \right) ^{3}}{(1+2\alpha \nu ^{1/3}+\nu ^{2/3}(\alpha ^{2}+1))^{3/2}}%
\ .  \label{superp}
\end{equation}%
Substituting the expression of $\tau \equiv \tan \chi $ as in (\ref{chisol}%
), one finds that
\begin{equation}
\frac{\left( 1-\nu \left( 1/{\tau }-\alpha \right) ^{3}\right) }{\left(
1-\nu ^{1/3}\left( 1/{\tau }-\alpha \right) \right) ^{3}}=\frac{1-\alpha \nu
^{1/3}+\nu ^{2/3}(\alpha ^{2}+1)}{1+2\alpha \nu ^{1/3}+\nu ^{2/3}(\alpha
^{2}+1)},
\end{equation}%
which yields the following explicit expression:%
\begin{eqnarray}
W &=&-\frac{1}{\sqrt{2}}q_{0}e^{-3\phi }\sqrt{1+2\alpha \nu ^{1/3}+\nu
^{2/3}(\alpha ^{2}+1)}\left( 1-\alpha \nu ^{1/3}+\nu ^{2/3}(\alpha
^{2}+1)\right) =\nn \\
&=&-\frac{1}{\sqrt{2}}e^{-3\phi }\sqrt{(q_{0}^{1/3}+(p^{0})^{1/3}\alpha
e^{2\phi })^{2}+e^{4\phi }(p^{0})^{2/3}}\ \ \cdot \nn \\
&&\qquad \qquad \cdot \left( (q_{0}^{1/3}+(p^{0})^{1/3}\alpha e^{2\phi
})^{2}+e^{4\phi }(p^{0})^{2/3}-3(q_{0}p^{0})^{1/3}\alpha e^{2\phi }\right) \
.  \label{superp-2}
\end{eqnarray}%
Notice that the overall minus in (\ref{superp-2}) is totally irrelevant,
since it can be eliminated with a $U(1)_{diag}$-rotation through the matrix
\ $-\epsilon\otimes id_{4}$\,.

Equation (\ref{superp-2}), up to a factor of $1/2$, coincides with the
formula of the non-BPS fake superpotential for the $(p^{0},q_{0})$
configuration in the $t^{3}$ model computed in \cite{Ceresole:2009iy}. The
difference of a factor $1/2$ is simply due to the different normalization
used for the normal form central charge in our notation (which coincides,
for example, with the one in Eq. (3.13) of \cite{Ferrara:2006em}) with
respect to the one used in \cite{Bossard:2009we}, as one can read from Eq.
(2.11) therein. This implies that the correct identification would be $%
\mathrm{Im}\mu _{0}=\frac{1}{2}\mathrm{Im}\zeta _{0}$. Consequently, the
correctly normalized fake superpotential becomes finally
\begin{eqnarray}
W &=&\frac{1}{2\sqrt{2}}e^{-3\phi }\sqrt{(q_{0}^{1/3}+(p^{0})^{1/3}\alpha
e^{2\phi })^{2}+e^{4\phi }(p^{0})^{2/3}}\ \ \cdot \nn \\
&&\qquad \qquad \cdot \left( (q_{0}^{1/3}+(p^{0})^{1/3}\alpha e^{2\phi
})^{2}+e^{4\phi }(p^{0})^{2/3}-3(q_{0}p^{0})^{1/3}\alpha e^{2\phi }\right) \
.  \label{superp-3}
\end{eqnarray}%
This computation is a non-trivial consistency check for the formalism based
on the axion-independent matrices $M$ and $\widehat{M}$ introduced in Secs. %
\ref{N=2} and \ref{Sec-2}, as well as for the results on the phase $\chi $
obtained above.


\section*{Acknowledgements}

We are glad to thank R. Kallosh for enlightening discussions and remarks. We
also acknowledge P. Aschieri and G. Bossard for useful discussions. The work
of A.C. and S.F. was supported by the European ERC Advanced Grant no. 226455
\textquotedblleft Supersymmetry, Quantum Gravity and Gauge Fields"
(SUPERFIELDS). The work of A.C. is also supported by the Italian MIUR-PRIN
contract 2009KHZKRX-007 \textquotedblleft Symmetries of the Universe and of
the Fundamental Interactions". The work of A.G. has been supported by the
Padova University Project CPDA105015/10.

\newpage

\begin{table}[h]
\begin{center}
\begin{tabular}{|c||c|c|c|c|}
\hline
$%
\begin{array}{c}
\\
J_{3}%
\end{array}%
$ & $%
\begin{array}{c}
\\
\frac{G_{4}}{H_{4}} \\
~~%
\end{array}%
$ & $%
\begin{array}{c}
\\
\frac{G_{5}}{H_{5}} \\
~~%
\end{array}%
$ & $%
\begin{array}{c}
\\
q \\
~~%
\end{array}%
$ & $%
\begin{array}{c}
\\
N \\
~~%
\end{array}%
$ \\ \hline\hline
$%
\begin{array}{c}
\\
J_{3}^{\mathbb{O}} \\
~%
\end{array}%
$ & $\frac{E_{7\left( 7\right) }~}{SU(8)}$ & $\frac{E_{6(6)}}{USp(8)}$ & $8$
& $8~$ \\ \hline
$%
\begin{array}{c}
\\
J_{3}^{\mathbb{O}_{s}} \\
~%
\end{array}%
$ & $\frac{E_{7(-25)}}{E_{6(-78)}\times U(1)}$ & $\frac{E_{6(-26)}}{%
F_{4(-52)}}$ & $8$ & $2$ \\ \hline
$%
\begin{array}{c}
\\
J_{3}^{\mathbb{H}} \\
~%
\end{array}%
$ & $\frac{SO^{\ast }(12)}{SU(6)\times U(1)}$ & $\frac{SU^{\ast }(6)}{USp(6)}
$ & $4$ & $2~$or$~6$ \\ \hline
$%
\begin{array}{c}
\\
J_{3}^{\mathbb{H}_{s}} \\
~%
\end{array}%
$ & $\frac{SO(6,6)}{SO(6)\times SO(6)}$ & $\frac{SL(6,\mathbb{R})}{SO(6)}$ &
$4$ & $0$ \\ \hline
$%
\begin{array}{c}
\\
J_{3}^{\mathbb{C}} \\
~%
\end{array}%
$ & $\frac{SU(3,3)}{SU(3)\times SU(3)\times U(1)}$ & $\frac{SL(3,\mathbb{C})%
}{SU(3)}$ & $2$ & $2$ \\ \hline
$%
\begin{array}{c}
\\
J_{3}^{\mathbb{C}_{s}} \\
~%
\end{array}%
$ & $\frac{SL(6,\mathbb{R})}{SO(6)}$ & $\left[ \frac{SL(3,\mathbb{R})}{SO(3)}%
\right] ^{2}$ & $2$ & $0$ \\ \hline
$%
\begin{array}{c}
\\
J_{3}^{\mathbb{R}} \\
~%
\end{array}%
$ & $\frac{Sp\left( 6,\mathbb{R}\right) }{SU(3)\times U(1)}$ & $\frac{SL(3,%
\mathbb{R})}{SO(3)}$ & $1$ & $2~$ \\ \hline
$%
\begin{array}{c}
\\
\mathbb{R} \\
(t^{3}\text{~model})~~%
\end{array}%
$ & $\frac{SL\left( 2,\mathbb{R}\right) }{U(1)}$ & $-$ & $-2/3$ & $2$ \\
\hline
$%
\begin{array}{c}
\\
\mathbb{R}\oplus \mathbf{\Gamma }_{m-1,n-1} \\
~%
\end{array}%
$ & {$\frac{SL\left( 2,\mathbb{R}\right) }{U(1)}\times \frac{SO(m,n)}{%
SO(m)\times SO(n)}$} & $SO(1,1)\times \frac{SO(m-1,n-1)}{SO(m-1)\times
SO(n-1)}$ & $\left( m+n-4\right) /3$ & $%
\begin{array}{c}
2~\left( m~\text{or~}n=2\right)  \\
4~(m~\text{or~}n=6) \\
~0~\text{otherwise}%
\end{array}%
$ \\ \hline
\end{tabular}%
\end{center}
\caption{Rank-$3$ Euclidean Jordan algebras $J_{3}$, and corresponding
symmetric scalar manifolds for vector multiplets in $D=4$ and $D=5$, with
the parameter $q$ and the number of supersymmetries ${N}$. }
\end{table}

\appendix

\section{\label{Res-Exp}Some Results on Exponential Matrices}

Let us recall the decomposition (\ref{AD}):%
\begin{equation}
\mathcal{A}=\left(
\begin{array}{cc}
\relax{\rm 1\kern-.35em 1} & 0 \\
\mathrm{Re}\mathcal{N} & \relax{\rm 1\kern-.35em 1}%
\end{array}%
\right) \left(
\begin{array}{cc|cc}
1 & 0 & 0 & 0 \\
a^{I} & 1 & 0 & 0 \\ \hline
0 & 0 & 1 & -a^{J} \\
0 & 0 & 0 & 1%
\end{array}%
\right) =(\mathcal{R})^{-1}\mathcal{A}_{D}(a^{I})\ ,
\end{equation}%
where $\mathcal{A}(a)=\exp (T(a))$ (\textit{cfr.} (\ref{def-1})).

Thus, by defining%
\begin{equation}
\mathcal{A}_{D}\equiv \exp (T_{D}),~\mathcal{R}\equiv \exp (T_{\mathcal{R}}),
\end{equation}%
and%
\begin{eqnarray}
T(a) &=&T_{D}(a)+T_{d}(a,d); \\
T_{D}(a) &\equiv &\left(
\begin{array}{cc|cc}
0 & 0 & 0 & 0 \\
a^{I} & 0 & 0 & 0 \\ \hline
0 & 0 & 0 & -a^{J} \\
0 & 0 & 0 & 0%
\end{array}%
\right) \ ,\quad T_{d}(a,d)\equiv \left(
\begin{array}{cc|cc}
0 & 0 & 0 & 0 \\
0 & 0 & 0 & 0 \\ \hline
0 & 0 & 0 & 0 \\
0 & 0 & d_{IJ} & 0%
\end{array}%
\right) ,
\end{eqnarray}%
one obtains that
\begin{equation}
\mathcal{A}(a)=\exp [T_{d}+T_{D}]=\exp [-T_{\mathcal{R}}]\cdot \exp [T_{D}]\
,
\end{equation}%
with
\begin{equation}
T_{\mathcal{R}}(d)\equiv \left(
\begin{array}{cc}
0 & 0 \\
-\mathrm{Re}\mathcal{N} & 0%
\end{array}%
\right) \ .
\end{equation}%
This allows us to describe how the matrix $\mathrm{Re}\mathcal{N}$ is
constructed from the algebra perspective, as
\begin{equation*}
\left(
\begin{array}{cc}
\relax{\rm 1\kern-.35em 1} & 0 \\
-\mathrm{Re}\mathcal{N} & \relax{\rm 1\kern-.35em 1}%
\end{array}%
\right) \equiv \mathcal{R}=\exp \left[ a^{I}(\hat{T}_{D})_{I}\right] \exp %
\left[ -a^{I}((\hat{T}_{D})_{I}+(\hat{T}_{d})_{I})\right] \ ,
\end{equation*}%
where the generators
\begin{equation*}
(\hat{T}_{D})_{I}=\frac{\partial }{\partial a^{I}}T_{D}\ ,\qquad (\hat{T}%
_{d})_{I}=\frac{\partial }{\partial a^{I}}T_{d}
\end{equation*}%
do not depend on the axions, since
\begin{equation*}
(\hat{T}_{{D}})_{I}\equiv \left(
\begin{array}{cc|cc}
0 & 0 & 0 & 0 \\
\delta _{I}^{J} & 0 & 0 & 0 \\ \hline
0 & 0 & 0 & -\delta _{I}^{J} \\
0 & 0 & 0 & 0%
\end{array}%
\right) \ ,\qquad (\hat{T}_{{d}})_{I}\equiv \left(
\begin{array}{cc|cc}
0 & 0 & 0 & 0 \\
0 & 0 & 0 & 0 \\ \hline
0 & 0 & 0 & 0 \\
0 & 0 & d_{IJK} & 0%
\end{array}%
\right) \ .
\end{equation*}


\section{\label{Expl-T^3}$M$ and $\mathcal{Y}$ in the $t^{3}$ model}

We now explicitly compute the matrices $M$ (\ref{MM-recast}) and $\mathcal{Y}
$ (\ref{YY}) for the special geometry defined by the holomorphic prepotential%
\begin{equation}
F=\frac{(X^{1})^{3}}{X^{0}},  \label{T^3}
\end{equation}%
corresponding to the $t^{3}$ model of $N=2$, $D=4$ supergravity, where the
unique complex scalar field is defined as%
\begin{equation}
\frac{X^{1}}{X^{0}}\equiv t=a-i\lambda .
\end{equation}%
In this model, which uplifts to $N=2$, $D=5$ \textquotedblleft pure"
supergravity (thus with no scalars in $D=5$), the matrices $M$ (\ref%
{MM-recast}) and $\mathcal{Y}$ (\ref{YY}) are simply numerical matrices.

From the analysis of \cite{CFM1}, it follows that
\begin{equation}
\partial _{i}K=6\lambda ^{2}\ ,\qquad g_{t\bar{t}}=12\lambda \ ,
\end{equation}%
with $\lambda =e^{2\phi }$. Since
\begin{equation}
a_{t\bar{t}}=\frac{1}{4}g_{t\bar{t}}e^{-4\phi }\ ,
\end{equation}%
it then follows that
\begin{equation}
(g^{1/2})_{t}^{\bar{t}}=2\sqrt{3}e^{2\phi }\ ,\qquad (a^{1/2})_{t}^{\bar{t}}=%
\sqrt{3}.  \label{formulae-1}
\end{equation}%
Thus, the matrices $M$ (\ref{MM-recast}) and $\mathcal{Y}$ (\ref{YY}) can be
computed to be
\begin{eqnarray}
M &=&\frac{1}{2}\left(
\begin{array}{cc}
1 & \sqrt{3}i \\
-\sqrt{3}i & -1%
\end{array}%
\right) =\text{sin}\left( \theta _{t^{3}}\right) \sigma _{3}\ +\text{cos}%
\left( \theta _{t^{3}}\right) \sigma _{2};  \label{M-T3} \\
\mathcal{Y} &=&\frac{1}{2}\left(
\begin{array}{cc|cc}
1 & 0 & 0 & -\sqrt{3} \\
0 & -1 & \sqrt{3} & 0 \\ \hline
0 & \sqrt{3} & 1 & 0 \\
-\sqrt{3} & 0 & 0 & -1%
\end{array}%
\right) ,  \label{Y-T^3}
\end{eqnarray}%
where $\sigma _{2}$ and $\sigma _{3}$ are the Pauli $\sigma $-matrices (such
that the constraints (\ref{UnitConstr}) are trivially satisfied), and%
\begin{equation}
\theta _{t^{3}}=\frac{\pi }{6},
\end{equation}%
such that (\textit{cfr.} (\ref{formulae-1}))%
\begin{equation}
(a^{-1/2})_{\overline{t}}^{t}=\text{tan}\left( \theta _{t^{3}}\right) .
\end{equation}

\section{\label{App-1}On the Complex \textit{Vielbein} for the $stu$
Parametrization of $N=8$ Supergravity}

The \textquotedblleft $stu$ parametrization" of $N=8$, $D=4$ supergravity is
based on the following correspondence between the skew-eigenvalues of the $%
N=8$ central charge matrix $Z_{AB}$ 
and the (flattened) scalar-dressed charges of the $N=2$, $D=4$ $stu$ model,
which is a common sector of all rank-$3$ symmetric special K\"{a}hler
geometries 
\cite{Ferrara:2006em,Ceresole:2009iy,Ceresole:2009vp}:
\begin{eqnarray}
Z_{AB} &=&\left(
\begin{array}{cccc}
z_{1}\epsilon & 0 & 0 & 0 \\
0 & z_{2}\epsilon & 0 & 0 \\
0 & 0 & z_{3}\epsilon & 0 \\
0 & 0 & 0 & z_{4}\epsilon%
\end{array}%
\right) =\nn  \label{ZAB} \\
&=&\left(
\begin{array}{cccc}
Z\epsilon & 0 & 0 & 0 \\
0 & -i(g^{s\bar{s}})^{1/2}\bar{D}_{\bar{s}}\bar{Z}\epsilon & 0 & 0 \\
0 & 0 & -i(g^{t\bar{t}})^{1/2}\bar{D}_{\bar{t}}\bar{Z}\epsilon & 0 \\
0 & 0 & 0 & -i(g^{u\bar{u}})^{1/2}\bar{D}_{\bar{u}}\bar{Z}\epsilon%
\end{array}%
\right) \ .
\end{eqnarray}%
The square root of $g_{i\bar{\jmath}}$ can in principle be chosen with real
entries as
\begin{equation}
(g_{s\bar{s}})^{1/2}=\pm \frac{i}{s-\bar{s}}\ ,  \label{choice}
\end{equation}%
and analogously for the $t\bar{t}$ and $u\bar{u}$ components of $g^{1/2}$.
Thus, in this symplectic frame, the rank-$3$ $C$-tensor reads
\begin{equation}
C_{stu}=\frac{i}{(s-\bar{s})(t-\bar{t})(u-\bar{u})}
\end{equation}%
can be written as
\begin{equation}
C_{stu}=\mp \ (g_{s\bar{s}})^{1/2}(g_{t\bar{t}})^{1/2}(g_{u\bar{u}})^{1/2}\ ,
\label{C-tensor}
\end{equation}%
consistent with the choice made in (\ref{choice}). This choice affects the
attractor equations since 
\begin{eqnarray}
\bar{Z}D_{t}Z &=&-iC_{stu}g^{s\bar{s}}g^{u\bar{u}}\bar{D}_{\bar{s}}\bar{Z}%
\bar{D}_{\bar{u}}\bar{Z}=\nn \\
&=&(\mp )(-i)(g_{t\bar{t}})^{1/2}(g^{s\bar{s}})^{1/2}(g^{u\bar{u}})^{1/2}%
\bar{D}_{\bar{s}}\bar{Z}\bar{D}_{\bar{u}}\bar{Z}\ ,\nn \\
&\Downarrow &\nn \\
Z(g^{t\bar{t}})^{1/2}\bar{D}_{\bar{t}}\bar{Z} &=&\mp i(g^{s\bar{s}%
})^{1/2}D_{s}Z\,(g^{u\bar{u}})^{1/2}D_{u}Z\ ,
\end{eqnarray}%
which, using the notations of (\ref{ZAB}), can be recast as
\begin{equation}
z_{1}z_{3}=\pm \overline{z_{2}}\overline{z_{4}}\ ,
\end{equation}%
where only the choice \textquotedblleft $-$" allows the attractor equation
from special geometry to be embedded into the $N=8$ theory. Thus, we are
lead to choose the minus sign in (\ref{choice}), and correspondingly the
\textit{Vielbein} is fixed to be purely imaginary:
\begin{equation}
\mathbf{e}=-i\mathbf{g}^{1/2}=\left(
\begin{array}{ccc}
({s-\bar{s}})^{-1} & 0 & 0 \\
0 & (t-\bar{t})^{-1} & 0 \\
0 & 0 & (u-\bar{u})^{-1}%
\end{array}%
\right) =-\overline{\mathbf{e}}.
\end{equation}

\section{\label{App-2}$U$-duality Invariants for the $D0-D6$ $i_{3}=0$
Configuration}

Following the definitions in \cite{CFMZ1,Ceresole:2009vp}, one can write the
following $U$-duality invariant expressions in $stu$ model within the $%
\left( p^{0},q_{0}\right) $ configuration with $i_{3}=0$ (recall (\ref{ZAB})
and (\ref{Zaxp0q0})):
\begin{eqnarray}
i_{1} &=&|Z|^{2}=2e^{-6\phi }q_{0}\left[ q_{0}+p^{0}\alpha _{1}\alpha
_{2}\alpha _{3}-e^{4\phi }p^{0}(\alpha _{1}+\alpha _{2}+\alpha _{3})\right]
\ ;\nn \\
i_{2}^{s} &=&|D_{s}Z|^{2}=2e^{-6\phi }q_{0}\left[ q_{0}+p^{0}\alpha
_{1}\alpha _{2}\alpha _{3}+e^{4\phi }p^{0}(-\alpha _{1}+\alpha _{2}+\alpha
_{3})\right] \ ;\nn \\
i_{2}^{t} &=&|D_{t}Z|^{2}=2e^{-6\phi }q_{0}\left[ q_{0}+p^{0}\alpha
_{1}\alpha _{2}\alpha _{3}+e^{4\phi }p^{0}(\alpha _{1}-\alpha _{2}+\alpha
_{3})\right] \ ;\nn \\
i_{2}^{u} &=&|D_{u}Z|^{2}=2e^{-6\phi }q_{0}\left[ q_{0}+p^{0}\alpha
_{1}\alpha _{2}\alpha _{3}+e^{4\phi }p^{0}(\alpha _{1}+\alpha _{2}-\alpha
_{3})\right] \ .
\end{eqnarray}%
It is worth remarking that that these four invariants collapse to a single
one, in the axionless case ($\alpha _{i}\equiv a^{i}/\lambda ^{i}=0$).

The black hole potential for this system is given in terms of the invariants
by
\begin{equation}
V_{BH}=i_{1}+i_{2}^{s}+i_{2}^{t}+i_{2}^{u}\ ,
\end{equation}%
and it admits the fake superpotential \cite%
{ADOT-1,Ceresole:2009iy,Bossard:2009we}
\begin{equation}
W=\frac{1}{2}\left( \sqrt{i_{1}}+\sqrt{i_{2}^{s}}+\sqrt{i_{2}^{t}}+\sqrt{%
i_{2}^{u}}\right) \ ;  \label{WW}
\end{equation}%
this case is usually referred to as the non-BPS \textquotedblleft
doubly-extremal" phase. Actually, one can show that (\ref{WW}) satisfies%
\begin{equation}
V_{BH}=W^{2}+4g^{i\bar{\jmath}}\partial _{i}W\overline{\partial }_{\bar{%
\jmath}}W  \label{rrel}
\end{equation}%
only in the case $i_{3}=0$. Indeed, by their very definitions, using the
special geometry relations (\textit{cfr. e.g. }Eqs. (2.24)-(2.26) of \cite%
{Ceresole:2009vp})
\begin{eqnarray}
D_{s}i_{1} &=&D_{s}i_{2}^{s}=\bar{Z}D_{s}Z\ ;\nn \\
D_{s}i_{2}^{t} &=&D_{s}i_{2}^{u}=iC_{stu}g^{t\bar{t}}g^{u\bar{u}}D_{\bar{t}}%
\bar{Z}D_{\bar{u}}\bar{Z}\ ,
\end{eqnarray}%
as well as the analogous ones concerning derivatives with respect to the
scalars $t$ and $u$, and by recalling that (recall (\ref{C-tensor}))
\begin{equation*}
C_{stu}{}^{2}=g_{s\bar{s}}\,g_{t\bar{t}}\,g_{u\bar{u}}\ ,
\end{equation*}%
one can compute that
\begin{eqnarray}
4D_{s}W\overline{D}_{\bar{s}}Wg^{s\bar{s}} &=&\frac{1}{4}\left[ (\sqrt{i_{1}}%
+\sqrt{i_{2}^{s}})^{2}+(\sqrt{i_{2}^{t}}+\sqrt{i_{2}^{u}})^{2}+\right. \nn \\
&&\phantom{\frac14}\left. +i\frac{(\sqrt{i_{1}}+\sqrt{i_{2}^{s}})(\sqrt{%
i_{2}^{t}}+\sqrt{i_{2}^{u}})}{\sqrt{i_{1}i_{2}^{s}i_{2}^{t}i_{2}^{u}}}%
(z_{1}z_{2}z_{3}z_{4}-\bar{z}_{1}\bar{z}_{2}\bar{z}_{3}\bar{z}_{4})\right] \
.
\end{eqnarray}%
By definition (\textit{cfr. e.g.} (1.12) of \cite{Ceresole:2009iy})%
\begin{equation}
i(z_{1}z_{2}z_{3}z_{4}-\bar{z}_{1}\bar{z}_{2}\bar{z}_{3}\bar{z}%
_{4})=i_{4}\Rightarrow i_{4}=-\sqrt{%
4i_{1}i_{2}^{s}i_{2}^{t}i_{2}^{u}-i_{3}^{2}},
\end{equation}%
thus
\begin{eqnarray}
V_{BH} &=&W^{2}+4g^{i\bar{\jmath}}\partial _{i}W\partial _{\bar{\jmath}}W+\nn
\\
&&-\left( \sqrt{i_{1}i_{2}^{s}}+\sqrt{i_{1}i_{2}^{t}}+\sqrt{i_{1}i_{2}^{u}}+%
\sqrt{i_{2}^{s}i_{2}^{t}}+\sqrt{i_{2}^{s}i_{2}^{u}}+\sqrt{i_{2}^{t}i_{2}^{u}}%
\right) \left( 1-\sqrt{1-\frac{i_{3}^{2}}{4i_{1}i_{2}^{s}i_{2}^{t}i_{2}^{u}}}%
\right) \ ,  \label{VV}
\end{eqnarray}%
which gives the required relation (\ref{rrel}) in the case $i_{3}=0$. We
also notice that the expression (\ref{VV}) is non-singular, since none of
the four invariants $i_{1},i_{2}^{s},i_{2}^{t},i_{2}^{u}$ vanishes for this
solution.

\end{document}